\documentclass[fleqn,usenatbib]{mnras}

\usepackage[shortlabels]{enumitem}
\setlist[enumerate, 1]{1\textsuperscript{o}}

\usepackage{newtxtext,newtxmath}

\usepackage[T1]{fontenc}

\DeclareRobustCommand{\VAN}[3]{#2}
\let\VANthebibliography\thebibliography
\def\thebibliography{\DeclareRobustCommand{\VAN}[3]{##3}\VANthebibliography}

\usepackage{graphicx}	
\usepackage{amsmath}
\usepackage{subcaption}
\usepackage{xcolor}
\usepackage{adjustbox}

\newcommand{\yatta}{YL~}
\newcommand{\teHS}{\theta_{\rm E}^{\rm S23b}}
\newcommand{\teAle}{\theta_{\rm E}^{\rm S19}}
\newcommand{\teOur}{\theta_{\rm E}^{\rm}}
\newcommand{\teYL}{\theta_{\rm E}^{\rm YL}}

\title[Lens model parameters with neural network]{Neural network prediction of model parameters for strong lensing samples from Hyper Suprime-Cam Survey}

\author[Priyanka Gawade et al.]{
Priyanka Gawade$^{1}$,\thanks{E-mail: priyankag@iucaa.in}
Anupreeta More$^{1,2}$,
Surhud More$^{1,2}$,
Akisato Kimura$^{3}$,
Alessandro Sonnenfeld$^{4,5}$,\newauthor
Masamune Oguri$^{6,7,2}$,
Naoki Yoshida$^{8,9,2}$
\\
\\
$^{1}$Inter-University Centre for Astronomy and Astrophysics, Ganeshkhind, Pune 411007, India.\\
$^{2}$Kavli Institute for the Physics and Mathematics of the Universe (WPI), University of Tokyo, 5-1-5,Kashiwa, Chiba 277-8583, Japan.\\
$^{3}$NTT Communication Science Laboratories, 3-1 Morinosato-Wakamiya, Atsugi Shi, 243-0198, Kanagawa, Japan.\\
$^{4}$Department of Astronomy, School of Physics and Astronomy, Shanghai Jiao Tong University, Shanghai 200240, China.\\
$^{5}$Leiden Observatory, Leiden University, P.O. Box 9513, 2300 RA Leiden, The Netherlands.
\\
$^{6}$Center for Frontier Science, Chiba University, Chiba 263-8522, Japan.\\
$^{7}$Department of Physics, Graduate School of Science, Chiba University, Chiba 263-8522, Japan.\\
$^{8}$Department of Physics, School of Science, The University of Tokyo, 7-3-1 Hongo, Bunkyo, Tokyo 113-0033, Japan.\\
$^{9}$Research Center for the Early Universe, School of Science, The University of Tokyo, 7-3-1 Hongo, Bunkyo, Tokyo 113-0033, Japan.\\
}

\date{Accepted XXX. Received YYY; in original form ZZZ}

\pubyear{2015}

\begin{document}
\label{firstpage}
\pagerange{\pageref{firstpage}--\pageref{lastpage}}
\maketitle

\begin{abstract}
Strong lensing of background galaxies provides important information about the matter distribution around lens galaxies. Traditional modelling of such strong lenses is both time and resource intensive. Fast and automated analysis methods are the need of the hour given large upcoming surveys. In this work, we build and train a simple convolutional neural network with an aim of rapidly predicting model parameters of gravitational lenses. We focus on the inference of the Einstein radius, and ellipticity components of the mass distribution. We train our network on a variety of simulated data with increasing degree of realism and compare its performance on simulated test data in a quantitative manner. We also model 182 gravitational lenses from the HSC survey using {\sc YattaLens} pipeline to infer their model parameters, which allow a benchmark to compare the predictions of the network. Given all considerations, we conclude that the network trained on simulated samples with lensed sources injected in empty HSC cutouts is the most robust, reproducing Einstein radii with an accuracy of about $10-20$ percent, a bias less than 5 percent, and an outlier fraction of the order of 10 percent. We argue in favour of the subtraction of the lens light before modelling the lens mass distribution. Our comparisons of the inferred parameters of 10 HSC lenses previously modelled in the literature, demonstrate agreement on the Einstein radius. However, the ellipticity components from the network as well as the individual modelling methods, seem to have systematic uncertainties beyond the quoted errors.
\end{abstract}


\section{Introduction}
\label{sec:intro}

Gravitational lensing can result in the formation of multiple images of sources which are sufficiently well aligned with a lensing potential. Observing and analysing such strong gravitational lensing effects, for example, the distortions of the background sources, the positions and magnifications of the multiple images, and the time delays in the case of transient sources, can provide us crucial information about the properties of the lens and the source. Strong lenses can be studied to probe the mass distribution in galaxies \citep[e.g.,][]{koopmans06, auger2010, sonnenfeld2015, shajib2021} and in groups and clusters \citep[e.g.,][]{limo2009, more2012cfhtls, oguri2012, newman2013, all2023}. 

Lens systems have been searched in ongoing imaging surveys such as the Subaru Hyper Suprime-Cam (HSC) survey through campaigns such as the Survey of Gravitationally-lensed Objects in HSC Imaging (\href{http://www-utap.phys.s.u-tokyo.ac.jp/~oguri/sugohi/}{SuGOHI}). This has resulted in the discovery of hundreds of strong gravitational lens candidates in various categories using different search algorithms \citep[see table 1 in][]{jaelani2023}. Lens systems have also been searched in surveys including DES \citep[e.g.,][]{ODonnell2022}, SLACS \citep[e.g.,][]{Bolton2006}, SL2S \citep[e.g.,][]{Sonnenfeld2013} and more recently, Euclid \citep[e.g.,][]{2025arXiv250209802N}.

The number of discovered gravitational lens systems is going to increase by an order of magnitude with the next generation ground based imaging surveys \citep[e.g.,][]{collet15} such as the Vera Rubin Observatory's Legacy Survey of Space and Time (LSST). Forward lens modelling techniques \citep[e.g.,][]{Night21....6.2825N, lenstronomy, SH10} are used to estimate the parameters that describe the mass distribution in such individual lens systems. Lens modelling is computationally expensive, often requires attention to individual systems, in addition to development of sophisticated lens modelling codes. Some of the steps that require human attention are identifying lensed features, masking out foreground contaminants, and finding an adequate initial guess for the lens model parameters. Finding more efficient ways to study gravitational lenses is the need of the hour given the large amount of data which will be available imminently. 

On the other hand, the era of large scale computing has already begun. Machine Learning (ML) has emerged as a revolutionary tool in order to mine data and identify features or various characteristics of the data \citep[e.g.,][]{review23}. The automation implies faster analysis and results. ML techniques such as neural networks are a very effective way in performing tasks like image classifications and parameter estimations. A neural network can be trained to analyse images of gravitationally lensed systems, for purposes like finding out strong lenses \citep[e.g.,][]{jacobs2017finding,rojas2022} and estimating lens model parameters \citep[e.g.,][]{hezaveh2017fast, Pearson19, Pearson21, HSIV, HSIX, S23b, lemon}. A neural network once trained, can predict the parameters just by analysing the features in the image such as shape, image separation and brightness.  A neural network can statistically infer the parameters of commonly used lensed models, with accuracy comparable to these sophisticated methods but about ten million times faster \citep{hezaveh2017fast}. Once we infer the lens parameters statistically, we can use them to probe various dark matter properties, for instance, the dark matter mass fraction \citep{oguri2014stellar} as a function of distance away from lensing galaxies and  mass distribution of lens systems \citep[e.g.,][]{more2012cfhtls}. 

\citet{hezaveh2017fast} demonstrated that they could use Convolutional Neural Network (CNNs) to infer the lens model parameters, in particular, the Einstein radius. Furthermore, \citet{levasseur2017uncertainties} have applied a Bayesian framework to determine the uncertainties in the lens parameters. These studies have mainly considered images of strong lenses at high spatial resolution, for instance, with image quality similar to that of the Hubble Space Telescope.

However, a significant fraction of strong lens systems in the near future will be obtained from the images taken by ground based telescopes, where we will have to deal with unique challenges including poorer image quality due to atmospheric seeing and the lower pixel resolution. To address this, we have designed a CNN that enables us to analyse the images of lens systems taken from ground based telescope surveys. A study of this nature has been carried out by \citet[][henceforth, referred to as P19]{Pearson19} focussing on simulations for lensed images detectable in ground based LSST as well as the space based Euclid mission. In our study, we focus on the HSC survey given the existence of data into which simulated lens systems could be injected, a sample of known lenses from the survey, and its similarity in terms of depth and image quality to the LSST. We train our CNN on a simulated sample of HSC-like lenses and test it on the simulated as well as real grade A and B SuGOHI lenses to infer lens mass model parameters like the Einstein radius, the axis ratio and the position angle of the major axis of the lens mass distribution. Unlike P19, who assumed perfect lens subtraction, we have to process the real SuGOHI lenses before feeding them to the network using  a pipeline called {\sc YattaLens} \citep[\yatta, ][]{2018son} in order to subtract the lens light and infer the lens model parameters to obtain the benchmarks for our study.

In \citet[][henceforth, referred to as S21]{HSIV}, they present a simple neural network to infer the lens model parameters of the simulated data, matched to the quality of ground based imaging data from the HSC survey. During the course of this work, \citet[][henceforth, referred to as S23a]{HSIX} presented a residual network (ResNet) for the same and in \citet[][henceforth, referred to as S23b]{S23b} they further applied the network on the real SuGOHI lenses.

Even though there are a number of similarities between our approaches, there are a number of differences in terms of the choice of the network, the suite and diversity of training samples, their weights and the test samples. In this study, we will also systematically analyse the reasons for the various failure modes of the network and show quantitative estimates of the performance of our network trained on a set of simulations with increasing amount of realism. Similar to S23b, we will also compare our results with the inferences from traditional modelling carried out in \citet[][henceforth, referred to as S19]{S19} and in S23b for a sample of 10 grade A SuGOHI lenses. Given these differences, our study is thus useful to establish the conclusions from P19 and S23b that are likely independent of the precise choice of network as well as the the impact of differences in training samples.

The goal of this paper is to enable fast modelling of strong lenses obtained from ground based surveys like the HSC using ML techniques such as CNNs and is structured as follows. In Section~\ref{sec:rconstruct_data}, we discuss the construction of simulated and real datasets that were used to train and test our CNN. Section~\ref{sec:nn_train} describes the architecture of our CNN and the training process in detail. In Section~\ref{sec:R_and_D} we discuss our results and we conclude with a summary in Section~\ref{sec:S_and_C}.

\section{Construction of training data}
\label{sec:rconstruct_data}

The performance of any machine learning architecture is dependent on the realism and representative nature on which the machine is trained. In this section, we describe the generation of the training and test data used to train and validate the network we construct. We also present details of the real lens samples from SuGOHI, including the pre-processing we performed before we used them as a test sample.

\subsection{Simulated datasets}
\label{sec:sim_data}
Gravitational lensing is a complex inverse problem. The inference of the parameters of the lens model has to be performed with the help of a noisy manifestation of the lensed source, where the true shape of the source is not known. Depending upon the unknown location of the source with respect to the lens as well as the lensing potential, the lensed images can take a variety of configurations. Therefore the CNN has to be trained on a large number of lensed images. The number of known galaxy-scale lenses from the HSC Survey are in $\mathcal{O}(100)$. Therefore, we generate a large number of lensed systems in order to train our network. We generate a sample of simulated lens systems using SIMCT \citep{simct}\footnote{https://github.com/anumore/simct}, a framework which allows for realistic lensed images to be generated for a given survey. Although SIMCT at its inception was used to generate lensed images from the Canada France Hawaii Telescope Legacy Survey (CFHTLS), we tweak its configuration to generate a final lensed image sample with properties that are matched to the HSC Survey depth, seeing and pixel resolution. While the detailed framework for SIMCT is given in \cite{simct}, we briefly summarise the methodology of producing the lens sample for completeness. 

The SIMCT pipeline generates lens samples via a hybrid approach where lensed features are model images superposed on real image cutouts of potential lensing galaxies with actual line-of-sight objects. The background source such as a galaxy or a quasar is defined with a parametric model and the parameter values come from realistic distributions of luminosity functions, redshifts, sizes and colors. The lens mass model is defined by taking the light properties such as the magnitudes, redshifts, ellipticity and position angle of the potential lensing galaxy and converting them into the parameters of a typical lens density profile such as the Singular Isothermal Ellipsoid (SIE) under the assumption that mass follows light and via standard scaling relations. External shear is drawn from a uniform distribution. Only those galaxies with sufficient lensing probabilities and lensed images that are detectable\footnote{We define detectability in this case by requiring that the image separation is greater than $0.5$ arcsecond, and the second brightest image is brighter than the magnitude threshold in the $r$-band.}, in a given survey data, comprise the final lens sample.

The simulated lensed arcs used in this work come from the same sample as that used in \cite{jaelani2023}. The reader is referred to Section 3 of \cite{jaelani2023} for the specific settings, the scaling relations and the distributions used to produce this sample. In Fig.~\ref{fig:sw}, we present the corresponding distributions of the lens mass model parameters under consideration. We note that we have added Poisson noise to the simulated model arcs after convolving with the coadded PSF from the HSC database
available at the location of central lens galaxy in each of the \textit{g}, \textit{r} and \textit{i} bands. 

We construct numerous training samples, successively, by adding various degrees of realism to our simulated arcs. These training samples were labelled as the following cases: a) Simulated arcs (PSF-convolved and Poisson noise added) without any background noise (henceforth, referred to as the PureSims sample), b) Simulated arcs, as before, but with Gaussian background noise corresponding to the HSC depth (henceforth, referred to as the GauNoise sample), c)Simulated arcs (same as in sample \textit{a}) added to selected cutouts from HSC with a size of 17~arcsec on the side. The cutouts are selected such that they do not contain any galaxy in the central region of radius of 5~arcsec and brighter than $24.5$ magnitude in the $i-$~band to avoid contamination of the lensed arcs. The images in this sample resemble real lenses after the ideal subtraction of the lens light (henceforth, referred to as the HSCempty sample). d) Simulated arcs (same as in sample \textit{a}) added to the image cutout of the respective HSC galaxy (henceforth, referred to as the LensLight sample). We would like to highlight that all of the above samples, except for the LensLight sample, contain images without the lens light. We will describe the purpose behind these different training samples while discussing the results in Section \ref{sec:R_and_D}.

\subsection{Real datasets}
\label{sec:real_data}
In addition to the sample of simulated lenses reserved for testing, we also make use of a real life test sample. This test sample of real lenses consists of 182 grade A and B \href{http://www-utap.phys.s.u-tokyo.ac.jp/~oguri/sugohi/}{SuGOHI} lenses. We are also using a sub-sample of 25 grade A lenses. These real images contain a lens galaxy at the center which in most of the cases outshines or contaminates the background lensed sources. It is mentioned in \cite{hezaveh2017fast}, where they use the high quality images of strong lenses from Hubble Space Telescope that the subtraction of the central lens light helped in improving their results. We are working with ground based data, where the blending is even more prominent. This prompted us to perform lens light subtraction on the SuGOHI lenses before feeding them to the network and we use the \yatta pipeline for this purpose. The \yatta pipeline fits the lens light and subtracts it out, identifies the lensed sources and then fits a SIE lens model to predict the parameters. We use lens-light subtracted images and further clean them by removing foreground objects identified by the \yatta segmentation results while constructing our test sample of real lenses. For grade A+B and grade A test samples, we use parameters predicted by \yatta to compare the predictions from our network. Finally, we find 10 SuGOHI lenses in the literature which have also been modelled by others with different algorithms for which we present a comparative analysis.

\section{Network Architecture and Training}
\label{sec:nn_train}
We develop a network following the conventional CNN architecture. Our network has seven convolutional blocks with alternate average pooling and two fully connected layers followed by an output layer. Each convolutional block has a convolutional layer followed by batch normalization, Parametric Rectified Linear Unit (PReLU) activation and a dropout layer with 20 per cent dropout rate. We use a mean-squared error (MSE) loss function and the Adam optimizer with a learning rate of 0.001 during training. Following the results of P19, we use data from multiple optical wavelengths and input the \textit{g}, \textit{r} and \textit{i} band images of a strong lens system. We use cutouts of $101\times101$ pixels for each band and expect the network to perform regression and output the lens mass model parameters we are most interested in, namely the Einstein radius, axis ratio and position angle. In the course of our investigations, we have also tried the addition of offsets between the lens galaxy light and lens potential of the magnitude similar to S21, and found not much of a difference in the network predictions for the parameters of our interest. This is likely due to the small magnitudes of these offsets. 

We train our model on a sample of 60000 simulated images of lens systems along with their lens model parameters namely Einstein radius, axis ratio and position angle. These 60000 images are obtained from augmenting 20000 unique images of strongly lensed systems by applying rotation. The distributions of the lens model parameters corresponding to these simulated lenses are obtained from the distributions of HSC lenses and mimic the real distribution of lenses in the Universe. These distributions are naturally imbalanced for certain parameters (see discussion in S21), and we will use explore the use of weights while we train our network. For instance, large Einstein radius systems are expected to be very few in number, which could result in poor training in the corresponding parameter range. At the same time, some of the parameters, such as ellipticity, are more difficult for a network to learn. Therefore, we have also explored the use of sample weights (i.e., weights corresponding to each training sample) and class weights (i.e., weights across the different parameters of a training sample). These weights modify the MSE loss during training, such that
\begin{align}
{\rm MSE} = \frac{1}{N}\left[ \sum_i w_i \sum_j \tilde{w}_j (\theta_j-\tilde{\theta}_j)^2 \right]
\end{align}
where $N$ is the total number of images in the batch, $w_i$ is the sample weight for the $i^{\rm{th}}$ image, $\tilde{w}_j$ is the class weight for the $j^{\rm{th}}$ parameter, $\theta_j$ represents true value of the $j^{\rm{th}}$parameter and $\tilde{\theta}_j$ corresponds to its value predicted by the network. We test for multiple values of sample weights and class weights monitoring the variance, the skewness and the kurtosis of the difference between the parameters predicted by our network and the underlying truth for our test samples. Based on these tests, we use sample weights equivalent to the square of the parameter value for the Einstein radius to counterbalance the lack of lenses with a higher Einstein radius (see Fig.~\ref{fig:sw}). We use class weights with values equal to 1, 10 and 5 for the Einstein radius, axis ratio and position angle, respectively.

We reserve 10 per cent of the training data for validation. We train our model with a batch size of 64. We apply an early stopping warning during the training in order to stop the training if the validation loss stops improving for 10 consecutive epochs. In our case as the validation loss starts plateauing, training stops after 40 epochs. We choose to monitor the validation loss instead of the training loss to avoid over fitting the model on the training data.

\begin{figure}
\begin{center} 
\includegraphics[scale=0.38]{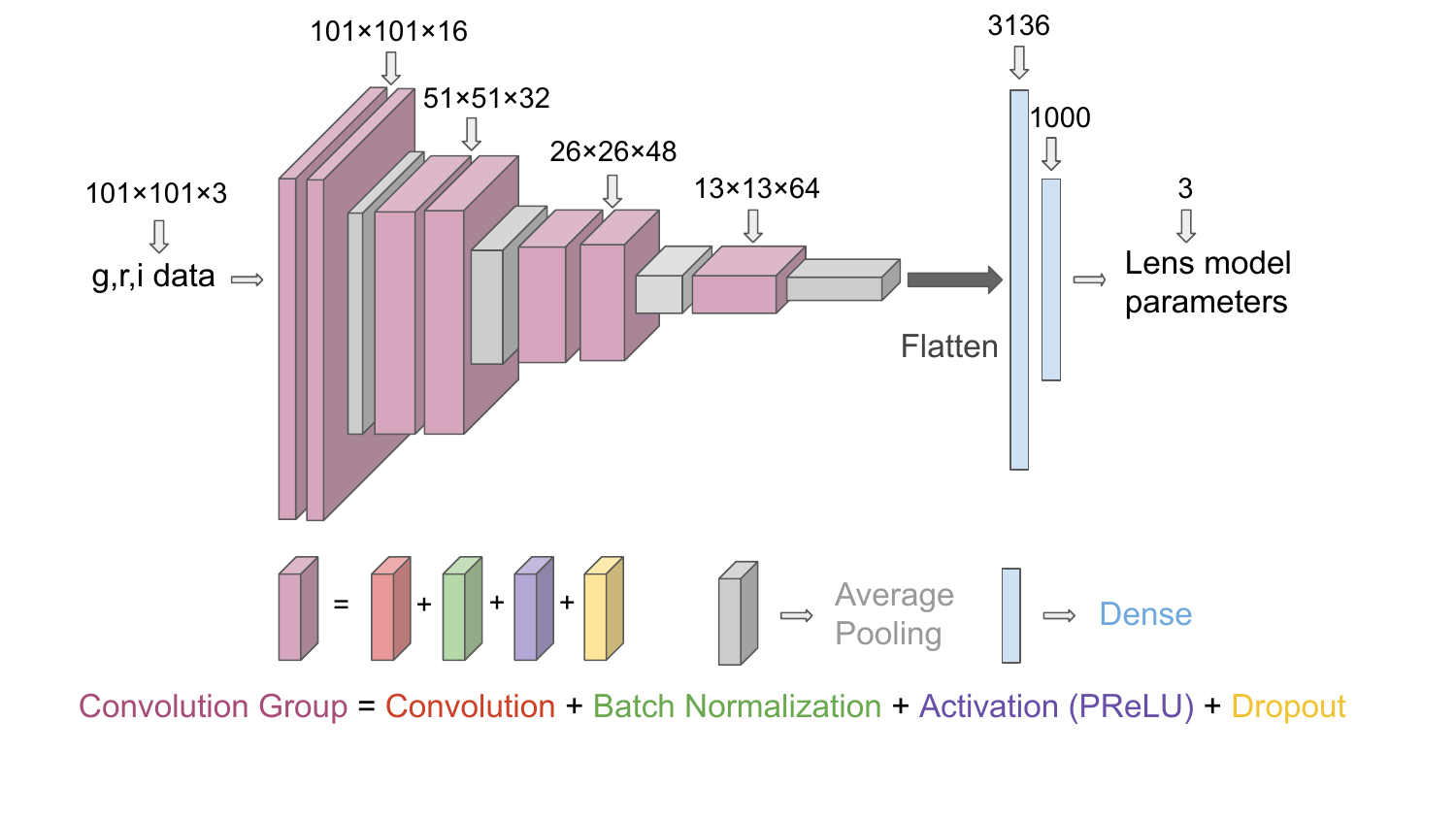} \caption{Network Architecture: The CNN that we use consist of a series of convolution groups interspersed with average pooling layers and terminating in two dense layers followed by an output layer as shown in the figure.}
\label{fig:net_arch}
\end{center}
\end{figure}

\begin{figure*}
\begin{center} 

\includegraphics[scale=0.325]{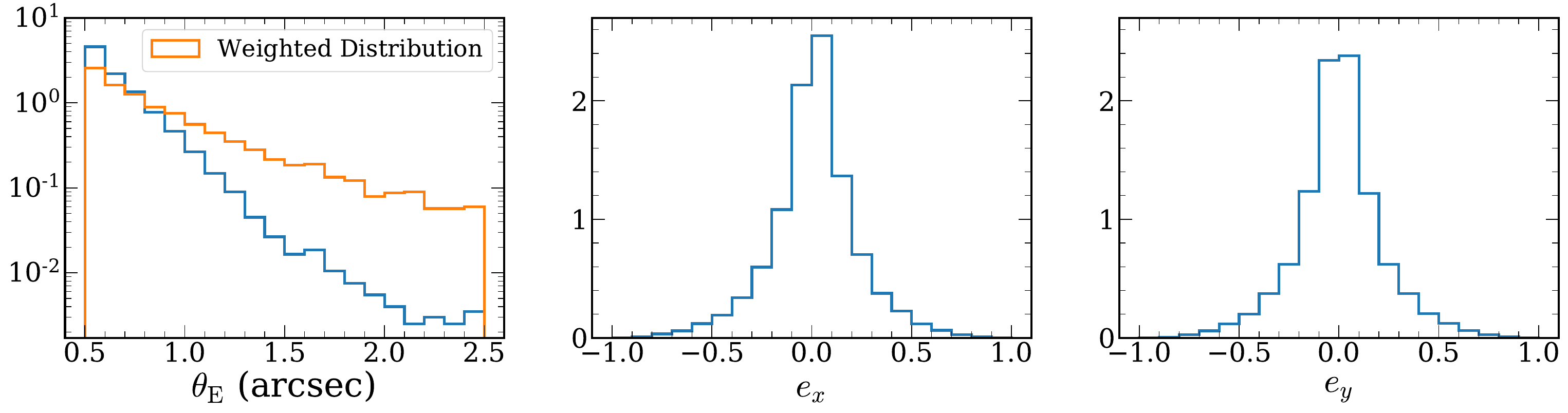}
\caption{The blue histograms show the normalized distributions of the parameters $\theta_{\rm E}$, $e_{x}$ and $e_{y}$ for the training sample. For $\theta_{\rm E}$, the orange histogram represents the distribution obtained after applying sample weights equivalent to $\theta_{\rm E}^2$, highlighting the effect of weighting to counterbalance the lack of lenses with higher Einstein radii.}
\label{fig:sw}
\end{center}
\end{figure*}

\begin{figure*}
\begin{center}
\includegraphics[scale=.45]{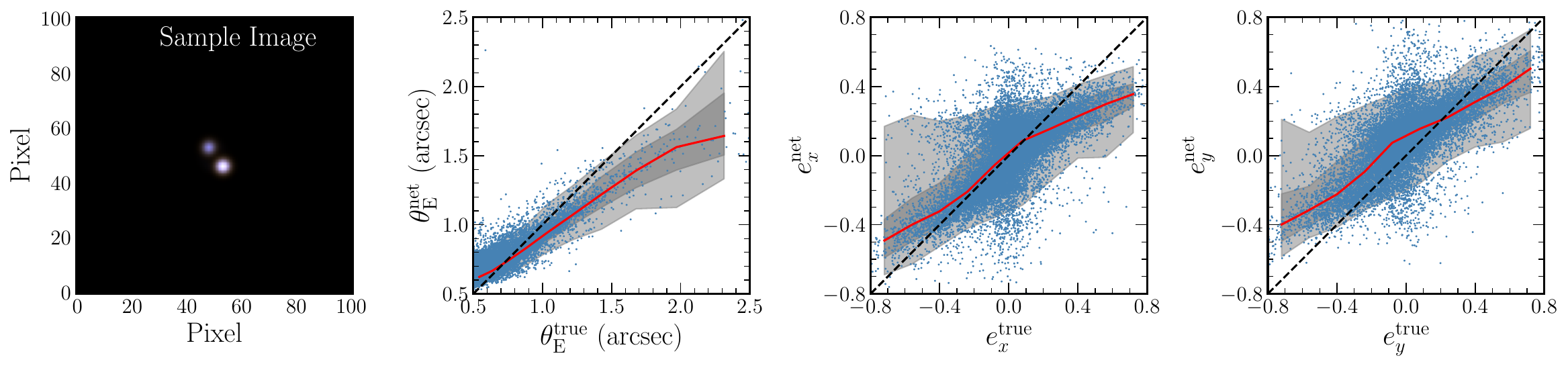}
\includegraphics[scale=.45]{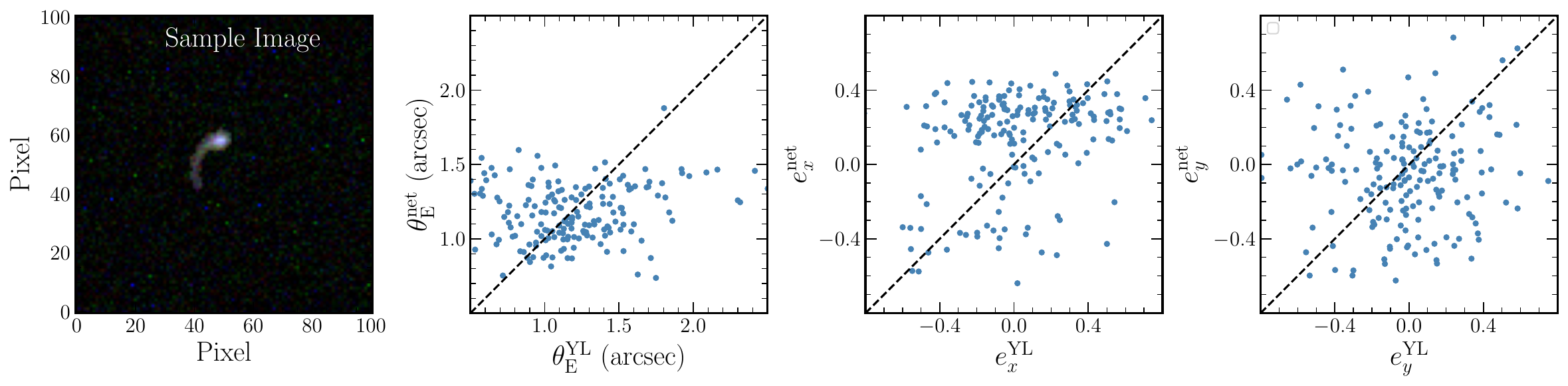}
\caption{Performance of our network trained with the PureSims sample (60000 images) on the corresponding test sample of 20000 lenses (upper row) and a test sample of 182 SuGOHI (grade A+B) lenses processed using \yatta to remove the central lens light and foreground objects (lower row). The size of the image cutouts (first column) is 101$\times$101 pixels corresponding to an angular size of $\sim~17$ arcsec $\times$ $17$ arcsec. The second, third and fourth columns show the network predictions for Einstein radius ($\theta_{\rm E}$), ellipticity components ($e_{x}$ and $e_{y}$), respectively, as compared to the true values. The range for every parameter is divided into 10 bins (logarithmic for $\theta_{\rm E}$ and linear for $e_{x}$ and $e_{y}$). The red curve represents the median and the dark gray and the light gray regions correspond to the 68 and 95 percentiles, respectively. 
}
\label{fig:PureSims}
\end{center} 
\end{figure*}

\begin{figure*}
\begin{center}
\includegraphics[scale=.45]{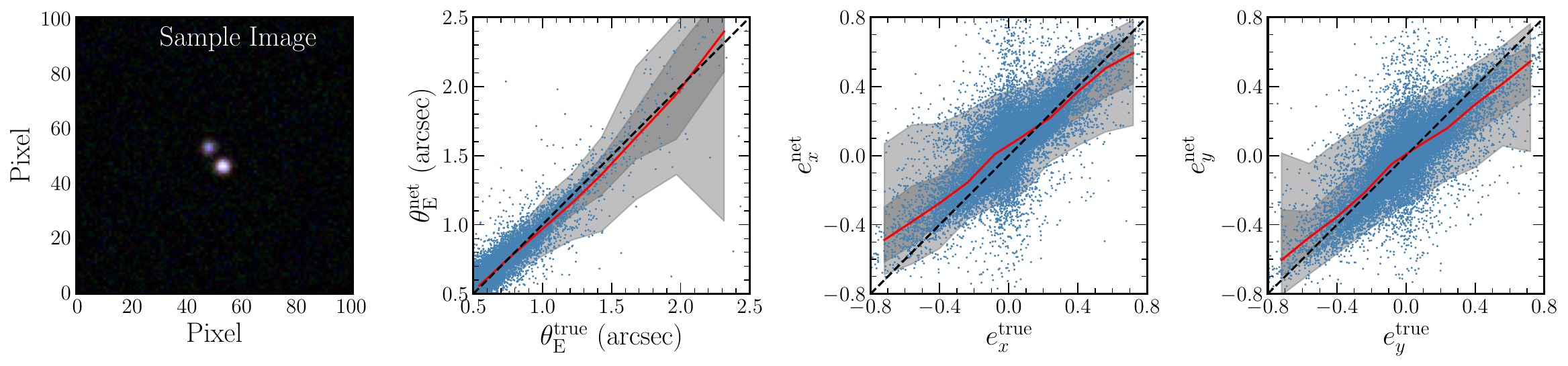}
\includegraphics[scale=.45]{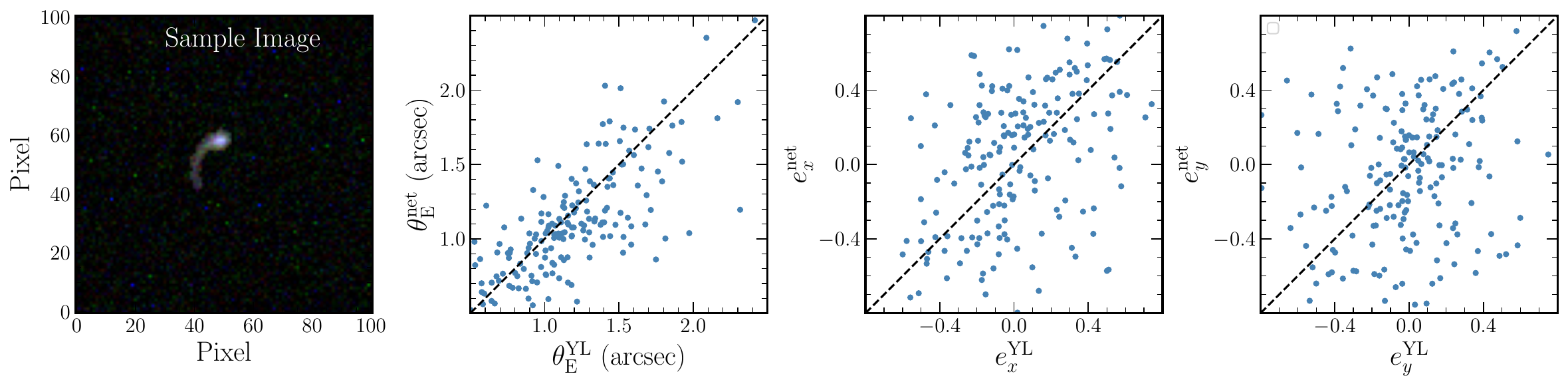}
\caption{Same as Fig.~\ref{fig:PureSims} except when the network is trained with the GauNoise sample.}\label{fig:GauNoise}
\end{center} 
\end{figure*}

\begin{figure*}
\begin{center}
\includegraphics[scale=.45]{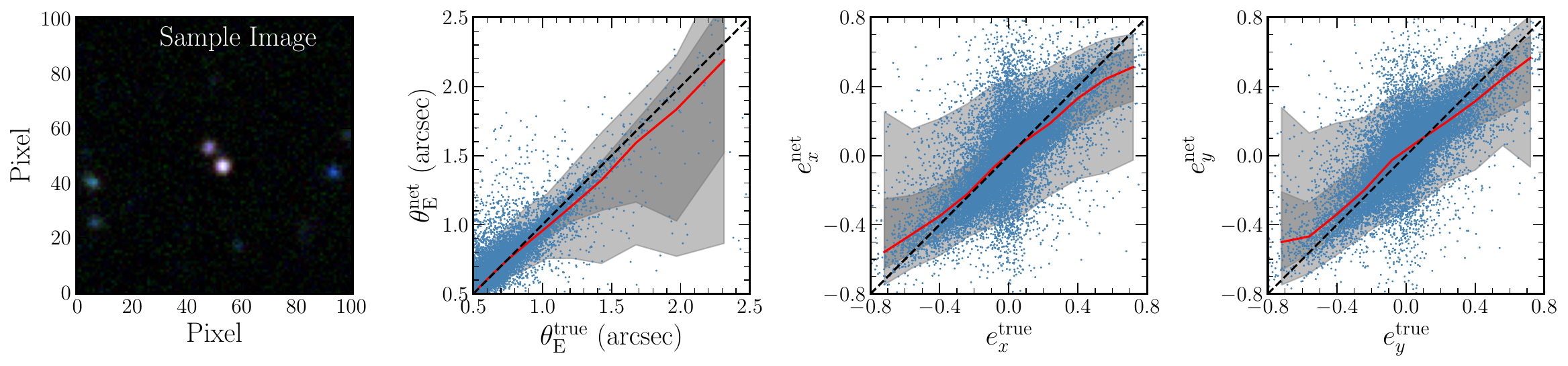} 
\includegraphics[scale=.45]{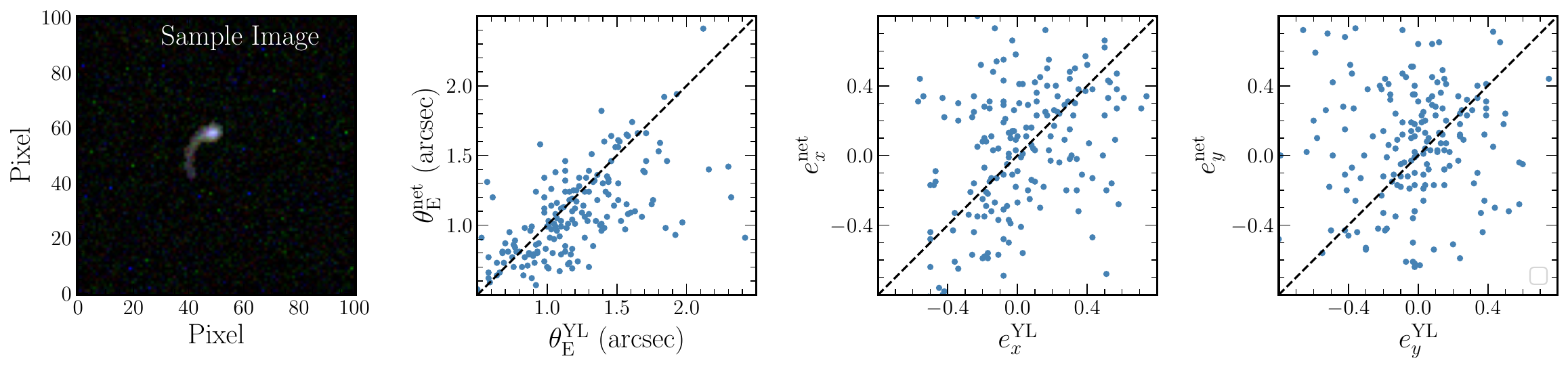}    
\caption{Same as Fig.~\ref{fig:PureSims} except when the network is trained with the HSCempty sample.}\label{fig:HSCempty}
\end{center} 
\end{figure*}

\begin{figure*}
\begin{center}
\includegraphics[scale=.45]{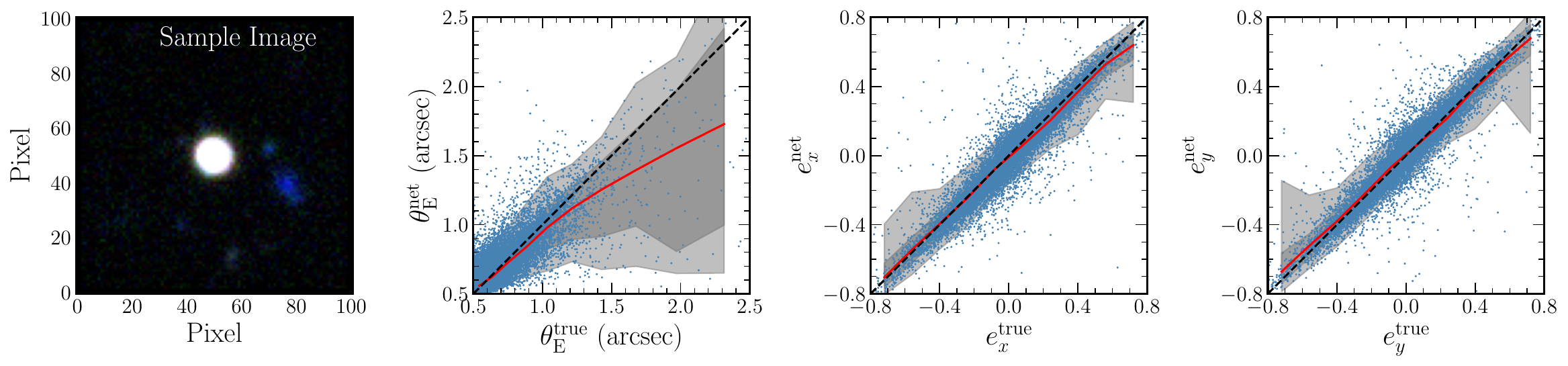}
\includegraphics[scale=.45]{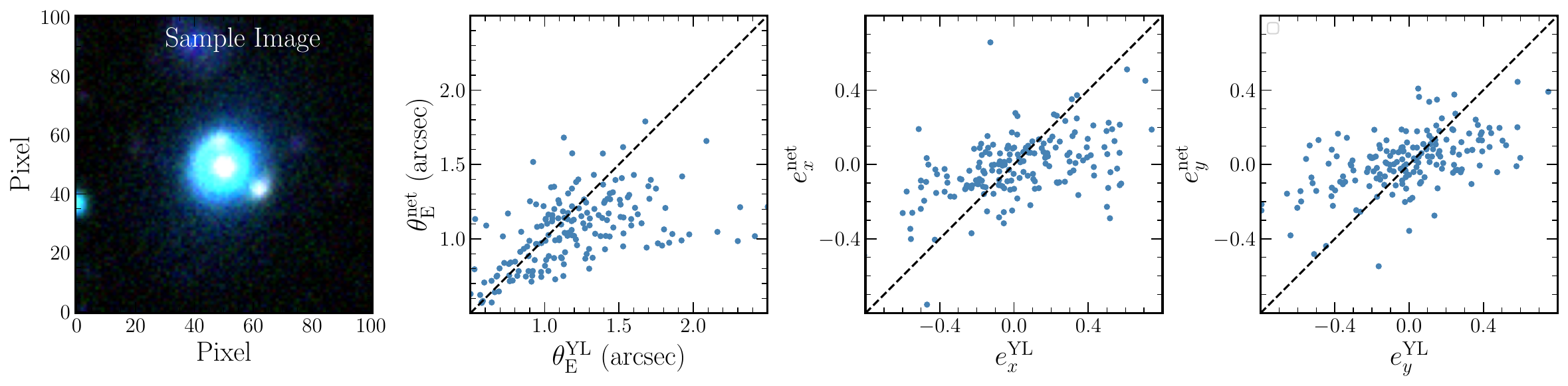}
\caption{Performance of our network when trained with the LensLight sample and tested on the corresponding simulated test sample (upper row) and the SuGOHI sample (not processed by \yatta, lower row).}\label{fig:LensLight}
\end{center} 
\end{figure*}

\begin{figure*}
\begin{center}
\includegraphics[scale=.45]{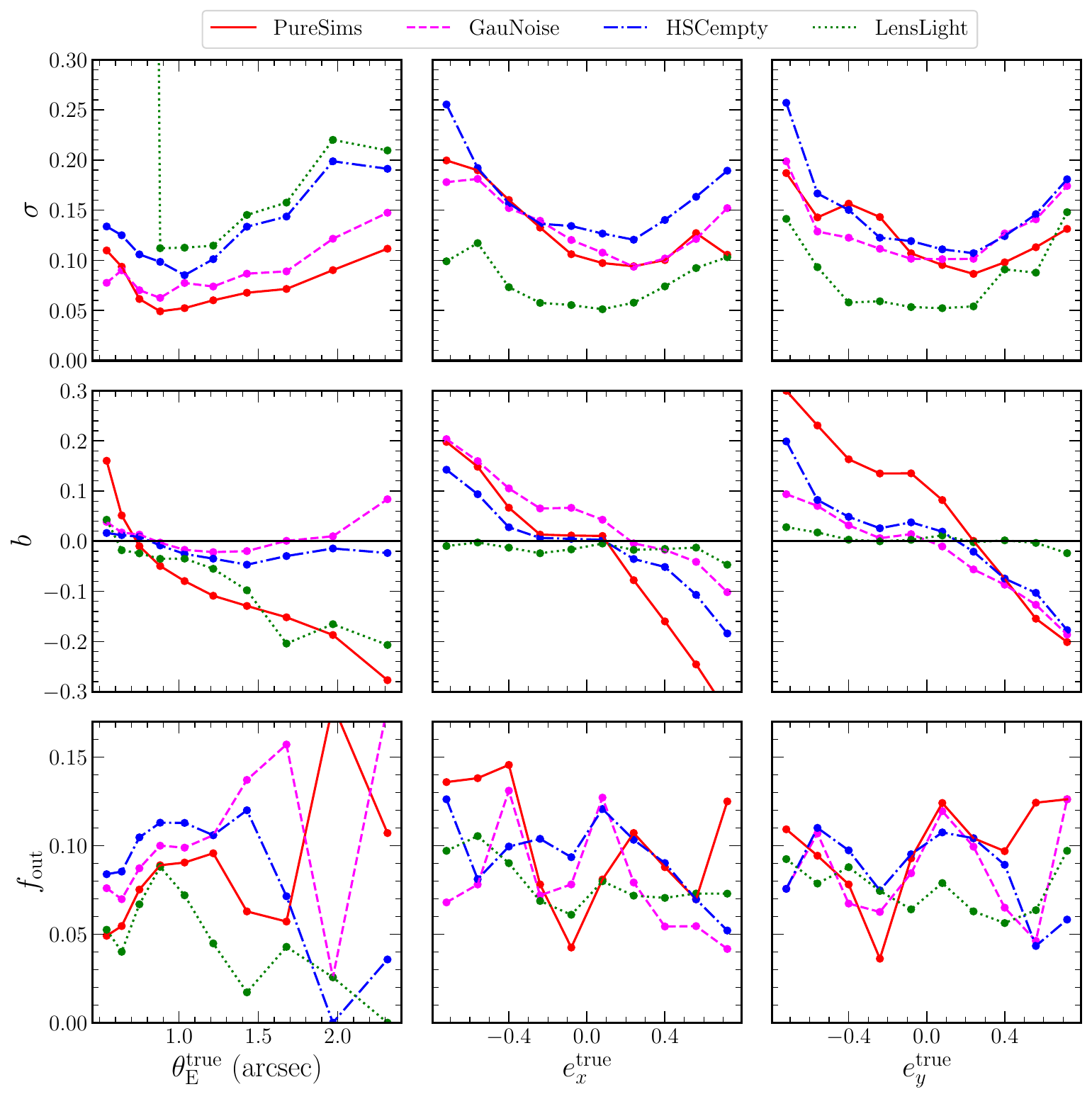}
\caption{Comparing the performance of our network statistically on four different samples, namely,  PureSims, GauNoise, HSCempty and LensLight represented by red solid, pink dashed, blue dashed dotted and green dotted lines respectively. The first, second and third column represent $\theta_{\rm E}$, $e_{x}$ and $e_{y}$ respectively. The first, second and third row represent the standard deviation ($\sigma$), bias ($b$) and outlier fraction ($f_{\rm {out}})$ respectively as defined in the beginning of the Section ~\ref{sec:R_and_D}.}\label{fig:Combat}
\end{center} 
\end{figure*}

\begin{figure*}
\begin{center}
\includegraphics[scale=.45]{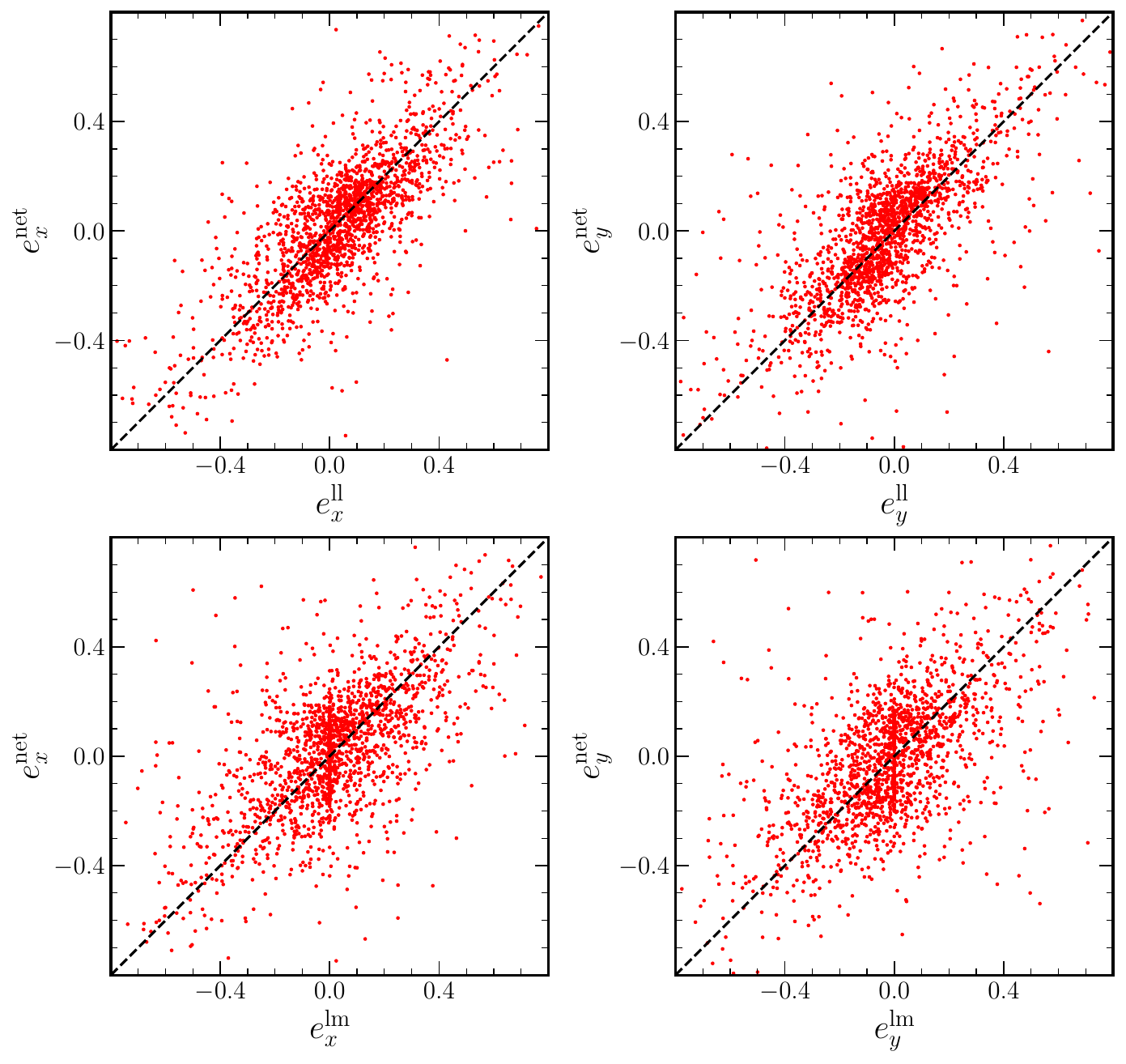}
\caption{Comparing the predictions of our network with true lens light ellipticity (upper row) and true lens mass ellipticity (lower row). In this case, we train our network on a training sample (similar to the LensLight sample) of 40000 lenses, with larger offsets between the lens light and the lens mass parameters ( $\sigma_{\rm q}=0.1$ and $\sigma_{\rm PA}=30$\textdegree) and test it on the corresponding test sample consisting of 1781 lenses. }
\label{fig:ll_lm}
\end{center} 
\end{figure*}

\begin{figure*}
\begin{center}
\includegraphics[scale=.45]{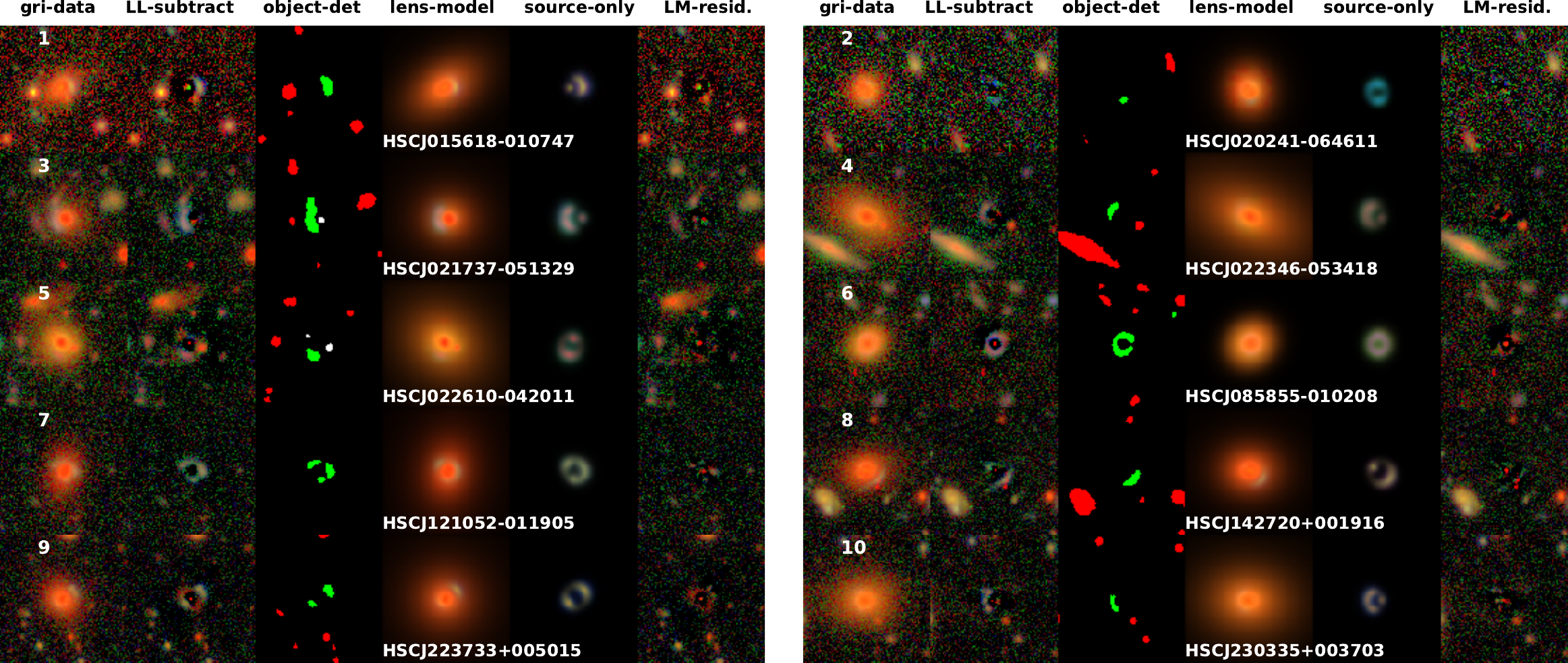}
\caption{A sample of colour-composite images of 10 SuGOHI lenses commonly modelled in S19, S23b, \yatta and our work. We have processed these lenses using \yatta in order to remove the central lens light and foreground objects. Columns from left to right : original image, lens light-subtracted image, arc and image segmentation map (green = arcs, white = modelled foreground objects, red = masked out foreground objects), best-fitting lens model of the system (including modelled foreground objects), best model of the lensed background galaxy alone (source only model), residuals between the data and the best-fitting lens model.}\label{fig:10_YL}
\end{center} 
\end{figure*}

\begin{figure*}
\begin{center}
\begin{subfigure}[b]{\textwidth}
    \includegraphics[width=0.49\textwidth]{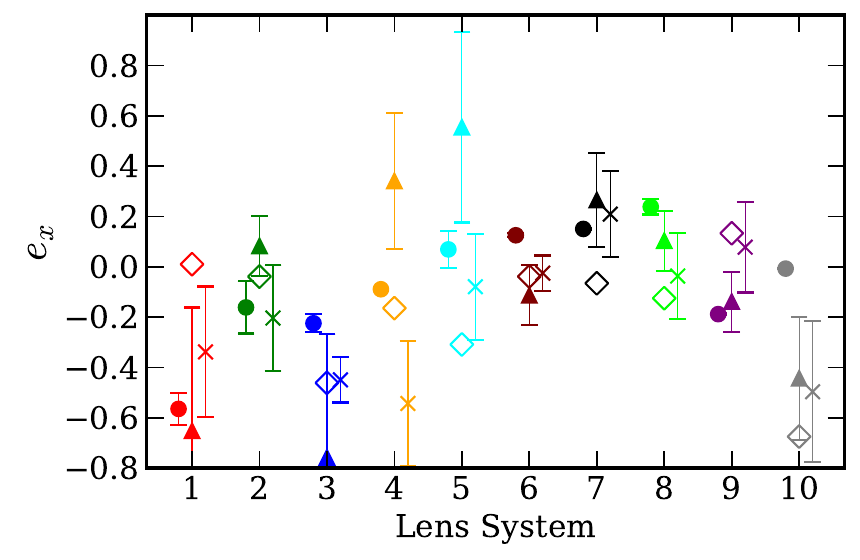} 
    \includegraphics [width=0.49\textwidth]{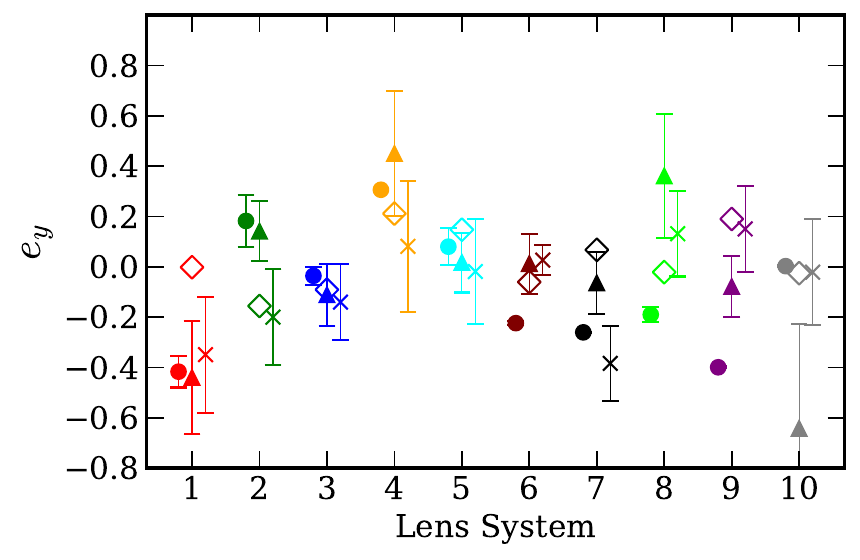}  
\end{subfigure}

\includegraphics[scale=0.625]{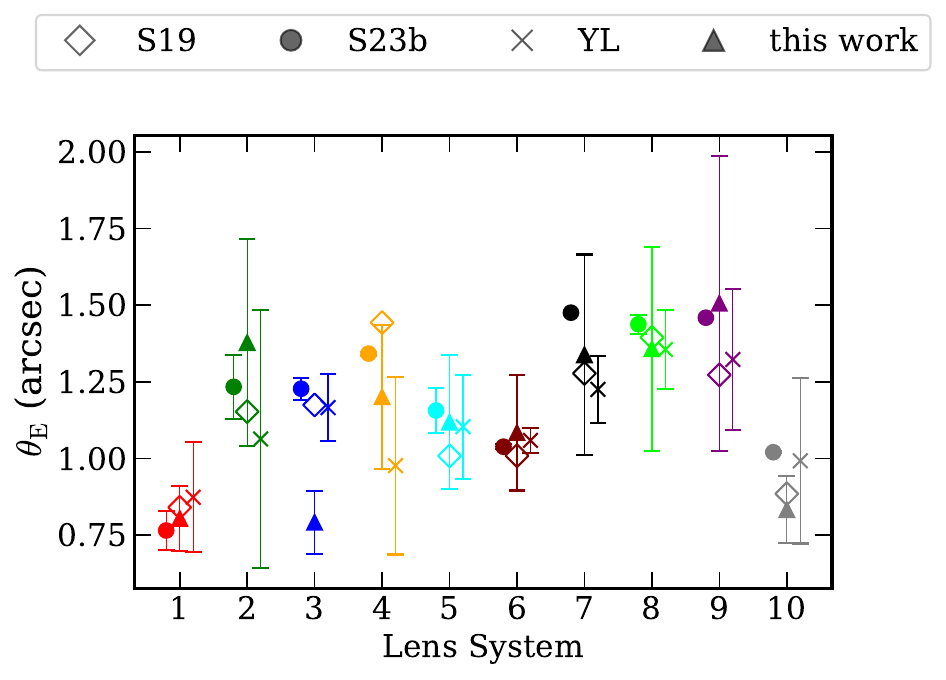} 
\caption{Comparison of the the lens model parameters, namely, Einstein radius ($\theta_{\rm E} $) and ellipticity components ($e_{x}$ and $e_{y}$ ), predicted by our network (this work) with \yatta, S19 and S23b results. The error bars on our network predictions are obtained as described in the Appendix ~\ref{app:errors}. We convert the results of \yatta, S19 and S23b to our conventions before comparing (see Appendix~\ref{app:conv}). In S23b, asymmetric uncertainties are reported for $\theta_{\rm E}$, $q$, and PA. We adopt the conservative bounds and propagate these errors assuming Gaussian distributions and treating the three parameters as independent. In contrast, S19 assumes constant uncertainties for each parameter ($\theta_{\rm E}$: 0.02 arcsec, $q$: 0.01, PA: 0.8°), which we omit from the plot for clarity.} 
\label{fig:10common}\end{center} 
\end{figure*}

\begin{table}
	\centering
	\caption{Lens model parameters predicted by our network (trained on the HSCempty sample) for 10 SuGOHI lenses that are also modelled in S19, S23b and our analysis with \yatta. The errors are obtained as described in the Appendix ~\ref{app:errors}.}
	\label{tab:10_table}
    \begin{adjustbox}{width=\columnwidth,keepaspectratio}
    \begin{tabular}{llccr p{10cm}}
        \hline
        ID  & Name & $\theta_{\rm E}$ (arcsec) & $e_{x}$ & $e_{y}$ \\
        \hline

1 &  HSCJ015618-010747 & $0.80 \pm 0.11$ & $-0.65 \pm 0.49$ & $-0.44 \pm 0.23$ \\
2 &  HSCJ020241-064611 & $1.38 \pm 0.34$ & $ 0.08 \pm 0.12$ & $ 0.14 \pm 0.12$ \\
3 &  HSCJ021737-051329 & $0.79 \pm 0.10$ & $-0.76 \pm 0.49$ & $-0.11 \pm 0.12$ \\
4 &  HSCJ022346-053418 & $1.20 \pm 0.24$ & $ 0.34 \pm 0.27$ & $ 0.45 \pm 0.25$ \\
5 &  HSCJ022610-042011 & $1.12 \pm 0.22$ & $ 0.55 \pm 0.38$ & $ 0.02 \pm 0.12$ \\
6 &  HSCJ085855-010208 & $1.08 \pm 0.19$ & $-0.11 \pm 0.12$ & $ 0.01 \pm 0.12$ \\
7 &  HSCJ121052-011905 & $1.34 \pm 0.33$ & $ 0.26 \pm 0.19$ & $-0.07 \pm 0.12$ \\
8 &  HSCJ142720+001916 & $1.36 \pm 0.33$ & $ 0.10 \pm 0.12$ & $ 0.36 \pm 0.25$ \\
9 &  HSCJ223733+005015 & $1.51 \pm 0.48$ & $-0.14 \pm 0.12$ & $-0.08 \pm 0.12$ \\
10 & HSCJ230335+003703 &$0.83 \pm 0.11$ & $-0.44 \pm 0.24$ & $-0.64 \pm 0.42$ \\
        \hline
    \end{tabular}
    \end{adjustbox}
\end{table}

\section{Results and Discussion}
\label{sec:R_and_D}
We train our network using simulated lenses and their true SIE lens model parameters, namely, the Einstein radius, axis ratio and position angle (see Appendix~\ref{app:conv} for conventions). We further convert the axis ratio and the position angle into the two components of the ellipticity ($e_{x}$ and $e_{y}$) as discussed in the Appendix~\ref{app:conv}. Amongst the three parameters Einstein radius is the most robustly predicted parameter as it depends on the radial angular separations of the lensed arcs from the central lens galaxy. However, the axis ratio and position angle are difficult to infer accurately. The presence of lens light may further worsen their prediction.

We train our network with four different simulated samples with increasing realism. The PureSims sample, where we convolve our simulated arcs with the PSF and add Poisson noise to them without adding any background noise, is a very ideal case. In this case we expect the network to be thrown off due to the presence of background galaxies, noise fluctuations in real images or imperfect subtraction of the lens light. In order to address these issues incrementally, we have used additional training samples. In the GauNoise training sample, we generate Gaussian background noise equivalent to our expectation in the HSC survey and add it to the PureSims sample, such that the network learns about the presence of noise and becomes insensitive to noise peaks. The HSCempty training sample, comprises of lensed source images added to cutouts from real HSC data which do not contain bright objects near the centre, and thus contains both noise and unassociated galaxies in the cutout. Finally, we also create a LensLight training sample, where the simulated lensed sources are added to the HSC image of the galaxy which was used to model the gravitational lensed images, in order to test whether lens light can further provide useful information for parameter estimation.

We test the performance of the network trained on each of these sample on the corresponding simulated test lens sample as we have true model parameters for comparing the network predictions and we also have statistically large lens samples to work with. The results of this exercise are shown in the upper panels of Figs.~\ref{fig:PureSims}-~\ref{fig:LensLight}. We quantify the predictions corresponding to these simulated test samples in a statistical manner. We divide the range of true values of each parameter in 10 bins, logarithmically for $\theta_{\rm E}$ and linearly for $e_{x}$ and $e_{y}$. 

In the case of the Einstein radius, $\theta_{\rm E}$, we compute the relative difference, $\theta_{\rm E}^{\rm{net}}/{\theta_{\rm E}^{\rm{true}}}-1$ and calculate the standard deviation and the median of this quantity as the scatter $\sigma$ and bias $b$ for $\theta_{\rm E}$, respectively. The scatter $\sigma$ and bias $b$ for the ellipticity components $e_{x}$ and $e_{y}$ is computed in a similar manner but using the difference between the network and the truth, ($e_{x}^{\rm{net}}-e_{x}^{\rm{true}}$ and $e_{y}^{\rm{net}}-e_{y}^{\rm{true}}$). We also devise an estimate of the outlier fraction ($f_{\rm{out}}$) for all the lens mass parameters. For a given simulated test sample, from amongst the top panels in Figs.~\ref{fig:PureSims}-~\ref{fig:LensLight}, we use the dark gray region corresponding to the 1$\sigma$ limit, and count the fraction of points that lie outside a region twice the 1$\sigma$ around the median for every bin. If the scatter is Gaussian distributed, then we expect, approximately 5 percent of the points to be quantified as outliers according to this definition. Any deviation of the outlier fraction, e.g. above (below) this 5 percent expectation, corresponds to larger (smaller) tails of the distribution.

The performance of our network on PureSims, GauNoise, HSCempty and LensLight test samples can be compared using the statistical quantities defined above and the standard deviation ($\sigma$), bias ($b$) and outlier fraction ($f_{\rm{out}}$) are shown in the three different rows of Fig.~\ref{fig:Combat}, respectively. The three columns correspond to the parameters $\theta_{\rm E}$, $e_{x}$ and $e_{y}$, respectively. 

For the Einstein radius, the PureSims trained network results in the lowest standard deviation in its predictions. However, this comes at the cost of an increasing bias as a function of increasing Einstein radius. The HSCempty sample has the lowest bias for the Einstein radius, although a larger scatter in the predictions. The outlier fraction estimates when trained on these network samples are quite noisy but seem to lie above the 5 percent thresholds in most cases, signifying the presence of non-Gaussian scatter in the predictions. The outlier fraction seems to be the lowest for the LensLight sample, again at the expense of an increasing bias at large Einstein radius. Overall the network trained on the GauNoise sample seems to represent a compromise in terms of the bias and scatter in the Einstein radius.

For the two ellipticity components, shown in the second and third column of the Fig.~\ref{fig:Combat}, respectively, the network trained on the LensLight sample results in the best results with the lowest scatter and bias, while the outlier fraction for all the samples are similar and hover around values between $5-12\%$. However, it is important to remember that the LensLight sample is simulated by keeping the components of ellipticity for the lens mass to be the same as the components of ellipticity for the lens light. Thus it is possible that the network is learning the components of ellipticity for the lens mass from the higher signal-to-noise ratio light from the lens present in the images, rather than from the lensed source configuration, itself. This also supports the presence of a larger bias and scatter in the Einstein radius prediction. 

In order to further find support for this scenario, we simulated samples similar to the LensLight sample but introducing an offset between the ellipticity components of the lens potential and the light distribution. First, we add a randomly drawn Gaussian offset with a scatter of $\sigma_{\rm q}=0.05$ and $\sigma_{\rm PA}=10$\textdegree for the axis ratio and the position angle of the lens light, respectively, to obtain the lens potential axis ratio and position angle. These offsets are identical to those used by S21. We further split each of these samples in training and test samples and analyze with our network. The predicted ellipticity components even in the case of these offsets are very tightly correlated with the lens potential ellipticity due to the small offset.

We further increased the width of the Gaussian offsets in the axis ratio and PA to $0.1$ and $30$\textdegree, to test the extent to which these conclusions are valid. In Fig.~\ref{fig:ll_lm}, we show the predicted ellipticity components with respect to the lens light and the lens potential in the top and bottom row panels for the second sample, respectively, for a network trained on the LensLight sample with increased offsets. We see that the network gives ellipticity components which are closer to the values of the light from the lens, than the lens potential. This shows that if the lens potential and the lens light are known to have fairly small offsets, then the network trained on LensLight can be quite useful to infer this ellipticity. However, in the presence of large offsets, the network may not retrieve the lens potential ellipticity components.

In addition to the simulated test samples, we also test our network trained on the different scenarios on the real SuGOHI lenses to assess the robustness of our network. When we train any CNN on a training sample which contains images having specific features, the network learns those features during the training process and we expect it to perform well when tested on the images having similar features. As a result, the performance of our network on the real SuGOHI lenses hugely depends on how well our simulated training images resemble the real SuGOHI lenses. 

The two main challenges we faced while carrying out this comparison is lens-light subtraction as well as the estimates of the true parameters of real lenses to compare our network predictions with. For this purpose, we use the \yatta pipeline with some minor changes suited for our analysis, to remove the lens light, identify lensed features, and run a traditional MCMC modelling pipeline \yatta on the SuGOHI lenses to obtain the lens model parameters and compare them with our network predictions.

A sample of SuGOHI lenses processed by \yatta is shown in the Fig.~\ref{fig:10_YL}. \yatta first models the light from the foreground lens galaxy in the centre and removes it to improve the identification of the lensed arcs in the background as shown in the second column (lens-subtracted) of Fig.~\ref{fig:10_YL}. It then distinguishes potential lensed arcs from the foreground objects as shown in the segmentation map in the third column (object detection) of Fig.~\ref{fig:10_YL}. \yatta then fits the SIE model to the lensed arcs in the lens light subtracted image as shown in the fourth column (lens model) of Fig.~\ref{fig:10_YL}. \yatta also estimates the lens model parameters which we have used to test our network.

The performance on the real SuGOHI lenses of our network when trained on the different samples is shown in the bottom panels of Figs.~\ref{fig:PureSims} - ~\ref{fig:LensLight}.  Even though the network trained on PureSims gave the least scatter on the simulated test sample, it fares very poorly on the real SuGOHI lenses, likely due to the missing features in the training such as background noise and the presence of other galaxies. In particular, the network prediction for the Einstein radius shows no correlation with the Einstein radius of the real lenses. We believe that this is likely a result of the network latching on to either noise or other unassociated galaxies in the image as being a part of the lensed images of the source or due to imperfect subtraction of the lens light from the real SuGOHI lenses.

The networks trained on more complex training samples help address these issues one by one. The performance on the real SuGOHI lenses of the network trained on the GauNoise sample is shown in the bottom panels of Fig.~\ref{fig:GauNoise}. Using a training sample which includes Gaussian noise improves the performance of the network on the real SuGOHI test images, where a positive correlation emerges in the network predictions of the Einstein radius compared to those from \yatta.

The performance on the real SuGOHI lenses, for a network trained on the Lenslight sample, is shown in the bottom panel of Fig.~\ref{fig:LensLight}. When compared to the results from the GauNoise sample, we see lesser correlation in the prediction of the Einstein radius on the real SuGOHI sample, and an almost flat prediction of the ellipticity components. Although this sample resembles well with the realistic scenario, as it has the lens light, the noise as well as other unassociated galaxies, 
it appears that the network is unable to adequately separate the light coming from the central lens galaxy, which either outshines or contaminates the background lensed arcs due to blending. This shows the importance of accurate modelling and subtraction of the lens light before the images are fed to the network.

The results for the real SuGOHI lenses with the network trained using the HSCempty sample, which has all the desirable features like HSC foregrounds and backgrounds but no lens light to contaminate the arcs, are shown in the bottom panel of Fig.~\ref{fig:HSCempty}. We present the corresponding lens model parameters predicted by the network and \yatta, along with their errors, in the Table~\ref{tab:182_table} (see Appendix~\ref{app:182AB} for details). Training on such samples improves the results for the SuGOHI test sample than when trained on the Lenslight sample and a performance comparable to the GauNoise sample.

Finally, we have also carried out cross-tests of the network trained on GauNoise sample with test lenses from HSCempty, and vice versa. Given that real lens galaxies are expected to have features such as background galaxies and real noise, we believe this is yet another important test, since the identification of lensed features by an alogorithm like \yatta may be susceptible to some failure modes. We find that the network trained on GauNoise performs significantly worse on all input parameters when tested on HSCempty sample, while the network trained on HSCempty sample performs robustly even on the GauNoise test sample.

Given all these consideration, we choose the network trained on the HSCempty sample to be our best case scenario. Although the GauNoise sample performs at par with the HSCempty sample on the simulated as well as SuGOHI lenses, it is important to consider the robustness of the network trained on the HSCempty sample. The fact that the HSCempty sample perform comparably with GauNoise sample indicates that the network trained on the HSCempty sample has also learned the real noise and the foreground objects and can distinguish them from the lensed arcs while predicting the lens mass parameters, which is important during application to real lenses.

We acknowledge that when we process the SuGOHI lenses using \yatta to remove the central lens light, sometimes the pipeline does not perform well either leaving a residue from the central lens light or modelling a part of the lensed source as a lens light and then subtracting it. As a result the lens light subtraction process can leave residuals which contaminate the actual configuration of lensed arcs, which can affect the predictions of the lens model parameters by \yatta. Given these issues, one may question the reliability of the parameters obtained by \yatta in order to test the network. Therefore, we also judge the performance of our network on a subset of real SuGOHI lenses which have been analysed previously by S19 and S23b and present this comparison in the next subsection.

\subsection{Network performance comparison of 10 real SuGOHI lenses commonly modelled by S19 and S23b}
\label{sec:s19_S23b}

A sample of 23 strong lenses from the constant mass (CMASS) sample of  Baryon Oscillation Spectroscopic Survey (BOSS) galaxies was modeled in S19 for which HSC imaging data in \textit{g}, \textit{r}, \textit{i}, \textit{z} and \textit{y} bands was available. They model the lens mass with a SIE profile running a MCMC code, using the software EMCEE \citep{MCMC}.  S23b applied a CNN to 31 grade A real galaxy-scale lenses from SuGOHI and compared their results with traditional, MCMC sampling-based models obtained from their pipelines GLEE \& GLAD. They also compared the results obtained from GLEE and GLAD with the results presented in S19 for some lenses common in both the analysis as shown in the fig.3 of S23b. We select 10 grade A SuGOHI lenses (see Fig.~\ref{fig:10_YL} and Table~\ref{tab:10_table}) that were commonly modelled by S19 and S23b.

While both of these methods agree well on the predictions of Einstein radius, they often differ significantly for the components of ellipticity (often beyond the quoted uncertainties). The fact that S19 uses SIE-only model, where GLEE and GLAD uses SIE+external shear model, alone is not enough to justify the discrepancy (see discussion in S23b).

We process this sample of 10 grade A SuGOHI lenses using \yatta as shown in Fig.~\ref{fig:10_YL} in order to remove the central lens light and foreground objects in order to feed it to our network that was trained on the HSCempty sample. We compare predictions from our network with the traditional MCMC modelling results presented in S19 and S23b in Fig.~\ref{fig:10common}. We convert the Einstein radius and position angle from the other two methods to our convention. We further convert the axis ratio and the position angle into components of ellipticity before comparing (see Appendix~\ref{app:conv}).

Here, we first describe our inferences from the visual comparison of the lens models shown in S19 (see fig.~1 in S19), S23b (see fig.~B.1 - fig.~B.31 in S23b) and \yatta (see Fig.~\ref{fig:10_YL})\footnote{Note: Since our network does not make predictions for all of the lens and source parameters yet, we cannot produce the equivalent ``best-fit'' model images for visual comparison.}. Next, we give a quantitative comparison of the model parameters from this work ( see Fig.~\ref{fig:10common}) and the other studies. The errorbars mentioned in the following text and as shown in the Fig.~\ref{fig:10common} are 1$\sigma$ errorbars from our network predictions and are discussed in  Appendix~\ref{app:errors}. 

\begin{enumerate}[1)]
     \item \textbf {HSCJ015618-010747} : In this case, the models of \yatta, S19 and S23b (see their fig.~B.1) are visually similar.
     The $e_{x}$ and $e_{y}$ values predicted by S23b, \yatta and network are within the quoted errorbars, while the predictions for $\theta_{\rm E}$ are similar for all the methods. The ellipticity components predicted by S19 are close to zero and deviate from other results.

     \item \textbf{HSCJ020241-064611} : For this lens, the models of \yatta, S19 and S23b (see their fig.~B.3) look qualitatively similar with one image in the north and the other in the south direction. The values of $\theta_{\rm E}$ for all four methods, are consistent within the quoted errorbars due to the similarity of models containing two compact images, whereas the ellipticity components are not.

     \item \textbf{HSCJ021737-051329} : The best-fit models of this lens by \yatta, S19 and S23b (see their fig.~B.5) are similar and agree well on $\theta_{\rm E}$ and $e_{y}$ . Our $\theta_{\rm E}$ and $e_{x}$ show deviation from other methods, while our $e_{y}$ agrees with them.

    \item \textbf{HSCJ022346-053418} : Since the counter image of the arc, if any, is barely visible in this lens, we expect that the degeneracies in various models will become more apparent here. The model images of \yatta, S19 and S23b (see their fig.~B.6) are visually similar. The model parameters $\theta_{\rm E}$ and $e_{y}$, for all four methods, are roughly consistent within the quoted errorbars, whereas the  $e_{x}$ values are poorly constrained.
     
     \item \textbf{HSCJ022610-042011}: In this case, the lens produces two images of the source galaxy. The model parameters $\theta_{\rm E}$ and $e_{y}$, for all four methods, are constrained well, while there is a huge inconsistency in the $e_{x}$ predictions.

      \item \textbf{HSCJ085855-010208} : The lens subtracted image in this lens system shows a near Einstein ring which is typical for a system with axial symmetry and a perfect alignment of the source with the lens center. The radius of the near perfect Einstein ring further helps each of the models to predict the parameter $\theta_{\rm E}$ accurately. The ellipticity components are also well constrained for S19, \yatta and our network, while predictions from S23b seem to deviate a from this. 
      
     \item \textbf{HSCJ121052-011905} : In this case, we can see the 3 images almost making an Einstein ring but two of them (the north-east and the west images) have extended structures deviating from the tangential direction. The models of \yatta and S19 are similar although the source of S19 is more compact than that of \yatta. These models lead to more smooth and circular configuration than the actual system. The S23b (see their fig.~B.15) source model is more clumpy and does not have extended features. The inferred $\theta_{\rm E}$ values for all the methods are within the errorbars because of the similar angular separations of the arcs from the center of the lens potential, whereas the ellipticity components are not well constrained due to the differences in the overall configuration.
     
     \item \textbf{HSCJ142720+001916} : For this lens, S19 and S23b (see their fig.~B.22) models appear more accurate with an extended source whereas \yatta model is more circular with a compact source. The models from all methods agree well on $\theta_{\rm E}$ and the predictions for $e_{x}$ are also within the errorbars, while the parameter $e_{y}$ is quite poorly constrained. 

      \item \textbf{HSCJ223733+005015} : In this case, \yatta and S19 models contain two diametrically opposite images and these models are qualitatively similar and accurate. As a result, these two methods also agree on the predictions of all the three parameters. The model of S23b (see their fig.~B.27) does not seem correct with an arc modelled incorrectly in the south-east direction. However, the results of S23b for $\theta_{\rm E}$ and $e_{x}$ agree well with our network.

      \item \textbf{HSCJ230335+003703} : For this quad lens system, S19 model looks quite accurate with distinct four images of a compact source. The lens model of \yatta does not seem accurate. It has an extended source and the model looks more circular than the actual configuration of the lensed images. The model of S23b (see their fig.~B.28) also does not do justice to the actual configuration. The $\theta_{\rm E}$ for all the methods is consistent within the quoted errorbars. The inferred $e_{y}$ from \yatta, S19 and S23b agree well but our network deviates from this, whereas for $e_{x}$ the predictions from \yatta and S19 are within the errorbars.
\end{enumerate}

After comparing results from the three conventional modelling methods, namely, \yatta, S19 and S23b with our network, we reach a similar conclusion to S23b. Even if sophisticated lens modelling methods are used on the same SuGOHI lenses with identical data quality, barring the Einstein radius, there is little consensus amongst the inferred parameters. The parameters of the ellipticity and its position angle, are likely heavily prior and or algorithm dependent. For instance, insufficiencies in modelling the light profile of the lens galaxy and the source, the choice of including or excluding the external shear and using different combinations of the broad-band data, can all heavily affect the answers beyond the quoted statistical uncertainties. In addition, the actual mass model of the galaxy can be more complicated than the SIE+external shear model. However, in the absence of the ground truth, one cannot assess which of the methods and their results are more accurate.
It may well be that the limitation is inherent to the quality and resolution of the ground-based survey data and better accuracy on the parameters is not possible unless working with data from space based surveys like the Hubble Space Telescope \citep[e.g.,][]{hezaveh2017fast}.

\subsection{Comparison of our network and results with studies from the literature}   
\label{comp_methods}

Our work has numerous similarities but also important differences to the growing literature where a neural network is used to analyse parameters of strong lensing systems, particularly from imaging surveys resembling the HSC survey (e.g., S21, S23a, S23b). We present a brief comparison of these methods in this section. The analysis done in P19 focuses on ground based imaging from the LSST survey, and given that the survey is yet to start, rely on training and testing their network on images where both the lens and the source images are simulated. The studies of S21, S23a, simulate source galaxies and inject them around real galaxies from the HSC survey. The study of S23b applies the network trained and developed in S23a to real SuGOHI lenses.

All the different groups carrying out these studies use simulation pipelines which are independently developed.  The simulations in P19 utilize cutouts with size $57$ $\times$ $57$ pixels corresponding to 11.4~arcsec on the side, for \textit{g}, \textit{r} and \textit{i} bands of LSST. They use a Sersic profile for both the source and lens galaxies and analyze images for cases including and excluding the lens light. In  S21, S23a, S23b, they simulate the strongly lensed systems in \textit{g}, \textit{r} \textit{i} and \textit{z} bands with a cutout size of $64$ pixels $\times$ $64$ pixels corresponding to 10.8~arcsec on the side, using HSC galaxies as lenses and HUDF galaxies as sources (keeping the lens light in their simulations). We use the SIMCT framework for our analysis and similarly use existing HSC galaxies as lenses but we keep the profiles of our source galaxies parametric, we only use \textit{g}, \textit{r} and \textit{i} bands with a larger cutout size of $101$ pixels $\times$ $101$ pixels corresponding to 17~arcsec on the side, while we exclude the lens light in the majority of our analysis runs separating out the problem of lens light subtraction and lens modeling analysis, although we do test the impact of including lens light in the analysis. However, for predictions of real SuGOHI lenses based on networks trained without lens light, we need to process these images in order to remove the lens light before feeding it to the network.

The network architecture and the training process play a crucial role in determining the accuracy of a neural network. Both P19 and S21 use a simple neural network, while in S23a and S23b they use a ResNet. Along with the differences in the network architectures, all of these studies differ from in the choice of lens model parameters they infer, loss functions, hyperparameters, sizes of the training, validation and test datasets and so on which could potentially lead to differences in their performance. The overall comparison of the results from our analysis using different simulated train samples and their respective test samples with the results from the aforementioned studies is as follows. 
 
P19 use LSST-like simulated data and use a simple neural network to estimate the Einstein radius, position angle and ellipticity of the lenses. When compared to our results on the aforementioned parameters with their LSST \textit{gri} case, their network seems to show results that are better than the network we present. We believe this excellent performance is related to the use of simulation for both lens and source galaxies considering a Sersic profile for the light. The performances of the networks when lenses include real HSC-lenses superposed on the simulated lensed arcs embedded in real HSC images paints a slightly different picture. The light profiles of real galaxies are often complex and any deviations from the simplicity could be interpreted as a  feature of the lenses source introducing systematic biases in the inferred parameters. This also may be part of the reason why the analysis done in P19 on simulated lenses with and without the lens light lead to only a slight degradation of the results. In comparison, our analysis of the SuGOHI lenses, especially for the Einstein radius parameter, with the network trained on the Lenslight sample (see the lower row of Fig.~\ref{fig:LensLight}) performs worse than the network trained on the HSCempty sample (the lower row of Fig.~\ref{fig:HSCempty}).

In S21, S23a and S23b, they simulate and analyse HSC-like images, similar to our analysis which enables a fairer comparison with these studies than P19. The analysis carried out in S21 using a simple CNN (see fig.~6 in S21) is quite similar to our analysis using the LensLight sample (see the upper row of Fig.~\ref{fig:LensLight}). Comparing these two plots, we infer that the predictions for the components of ellipticities from both the studies appear qualitatively similar. The use of lens light to determine the ellipticity works best as long as there is little scatter between the ellipticity of the lens light and the lens potential. For the Einstein radius, although S21 results appear slightly better than our results with the LensLight sample, they are comparable with our predictions for the Einstein radius with the HSCempty sample.

In S23a, the authors build upon the results of S21 with a ResNet which improves the recovery of any spatial offsets between the lens light and mass distribution. However, although the ResNet improved their results on the $x$ and $y$ dramatically (see fig.~4 in S23a) compared to S21, the results on the components of ellipticity remain comparable. The results in S23a, corresponding to the Einstein radius (see fig.~4 in S23a) seem to slightly degrade compared to S21. Although S23a additionally include external shear, the network has difficulty in the inference of the external shear. We find that the performance of our network is better or comparable to the case of S23a.
The performance of the network from S23a with traditional lens modelling methods on grade A SuGOHI sample of lenses was presented in S23b and can be compared with the comparison between our network and traditional lens modelling method (see Appendix ~\ref{app:gradeA}). We see that the networks in both these studies have comparable performance in reproducing the results from traditional modeling methods for both the Einstein radius and the two components of the ellipticity. 

Even though the studies compared in this section, all aim towards a common scientific goal, differences in the simulation methods, datasets, network architectures, lens model parameters and overall results make these studies interesting in their own right to build a consensus on the direction of required efforts. The discrepancy found in the results of modelling of lenses with traditional methods (see Fig.~\ref{fig:10common}) invites more efforts and independent investigations along these lines.

\section{Summary and Conclusion}
\label{sec:S_and_C}

The next generation imaging surveys are expected to increase the number of known strong lensing systems to $\mathcal{O}(10^5)$, which will present formidable challenges for modelling each lens system individually. Analysis of images of strong lens systems from a ground-based survey presents a challenge, as the imaging is limited by poor image quality due to atmospheric seeing and low angular resolution. Additionally, increasing depths of the surveys implies increased number of lenses at higher redshifts where the lensed images can often be faint. In this work, we systematically investigate the ability of neural networks to analyse the strong lenses from ground-based imaging surveys in a fast and automated way and its performance depending upon the realism of the training samples used.

We developed a simple CNN to analyse the strong lenses from the HSC data, a precursor to LSST, and estimate the parameters of the SIE lens mass model, namely, the Einstein radius, the axis ratio and the position angle of the major axis of the mass distribution. We first trained and tested our network on 60000 HSC-like galaxy-scale simulated lenses. We prepared four different training samples with increasing degree of realism: PureSims sample included only the arcs with Poisson noise, the GauNoise sample further included background Gaussian noise, the HSCempty sample included the lensed features on top of randomly selected cutouts from HSC with central ``empty" regions, and the LensLight sample also included the light from the lens galaxy. 

We trained our network on each of the four types of samples and checked its performance on both the corresponding simulated test samples and the real sample of 182 galaxy-scale SuGOHI lenses (grade A and B) to predict the lens parameters. The performance of the network on SuGOHI lenses was tested against the parameters predicted by \yatta for these systems, which were analysed for the first time in this paper.

We summarise the results of our the performance of the CNN that we developed and comparison with the literature studies below.

\begin{itemize}
    \item Across the investigations carried out in this work, we find that the Einstein radius of a lens system is the most robust parameter that can be inferred, both from traditional modelling as well as machine learning.
    
    \item The performance of the CNN that we have constructed varies from parameter to parameter depending upon the training sample used to train the network. We have shown that the network trained on the HSCempty simulated sample of lenses is the most robust. This network can predict the Einstein radius with a scatter of 10-20 percent, a bias less than 5 percent and an outlier fraction around 10 percent.
    
    \item On the other hand, the network trained on the Lenslight simulated sample predicts ellipticity parameters more accurately. We show, however, that the improved accuracy of the network trained on the Lenslight sample is correlated to the lens light parameters more than the lens mass parameters. If the real lens galaxies have larger misalignment between the lens mass and light parameters, then such networks are likely to have increased scatter. Therefore, for modelling inferences which are independent of the lens light, our results suggest that it is ideal for networks to be trained after lens light subtraction and that accurate lens light subtraction techniques be further investigated.
    
    \item We compared our work with similar studies in the literature, where a neural network is trained on HSC-like data to predict the lens mass parameters. In spite of using completely independent methodology for simulating lenses, differences in choices or assumptions when generating the lens populations, different network architectures and different means of estimating true (or reference) parameters for real lenses, our accuracies and scatter on the predicted lens mass parameters, especially the Einstein radius, are comparable.
    
    \item We have modelled the 182 SuGOHI (Grade A and B) lenses together for the first time using \yatta, and presented the parameters and their errors. When the  network predictions are compared to those from \yatta, we find that the network trained on HSCempty sample performs the best amongst the alternatives we explore.
    
    \item We further compared the predictions from our network, trained on the HSCempty sample, for 10 SuGOHI lenses that are also modelled by others (S19, S23b) and by us with \yatta using conventional MCMC modelling methods. We find that the traditional methods agree on the Einstein radius but the two components of ellipticity show larger scatter even when the lenses are modelled on a case-by-case basis using sophisticated modelling techniques (e.g. S19 and S23b). This may either due to the limitation of ground-based data quality, lens model degeneracies or lack of understanding of misalignment between mass and light components.

    \item Networks trained on all samples perform well on the simulated samples and somewhat poorly on the real samples. This holds true even in those networks where the simulated lenses include some misalignment between lens mass and light parameters (see results of S23b, which were found to be consistent with our results on the grade A SuGOHI sample) suggesting a) scope for improvement in the quality and diversity of simulated samples and b) possibly, need for development of a more robust network 
    
\end{itemize}  

Overall our network, once trained, takes around a millisecond to predict the lens mass parameters on a single lens-light subtracted image. This makes our analysis equipped for analysing a plethora of strong lenses $\mathcal{O}(10^5)$, expected from the next generation surveys like LSST in a reasonable amount of computational cost.

In spite of the ongoing progress in large-scale modelling of lenses, some challenges remain to be addressed hitherto. For instance, we note that processing the SuGOHI lenses via \yatta, to remove the central lens light, may introduce artifacts in the lensed images either by leaving a residue or over-subtracting the flux. Such artifacts can contribute to increased parameter uncertainties,  particularly, for small Einstein radius systems. The absence of true model parameters for the real lenses, and the non-consensus of parameters from the conventional modelling techniques make it harder to objectively quantify the network performance.

In future, we plan to incorporate prediction of the additional lens and source parameters along with analysing how the network performance varies as a function of SNR, number of detected images and presence of foreground contaminants. Furthermore, we also are exploring the using the interpretability tools for CNNs to understand the model predictions by our network and its failure modes. We aim to extend this work to upcoming ground based surveys like LSST.

\section*{Acknowledgements}
\label{sec:aknow}
We thank Stefan Schuldt, Francisco Villaescusa-Navarro, Vibhore Negi, Shreejit Jadhav, Vishal Upendran and Navin Chaurasiya along with Sukanta Bose for useful discussions on the project. We also thank Stefan Schuldt for sharing the data. PG acknowledges financial support provided by the University Grants Commission (UGC) of India.  We acknowledge the use of the high performance computing facility Pegasus at IUCAA for this work. NY and AK thank financial support by Japan Science and Technology Agency AIP Acceleration Research Grant Number JP20317829 and JSPS Kakenhi 24H00221.

\section*{Data Availability}
\label{sec:data_avail}

The {\href{http://www-utap.phys.s.u-tokyo.ac.jp/~oguri/sugohi/}{SuGOHI}} \footnote{\url{http://www-utap.phys.s.u-tokyo.ac.jp/~oguri/sugohi/}} lenses and the {\href{https://github.com/astrosonnen/YattaLens}{\yatta}}\footnote{\url{https://github.com/astrosonnen/YattaLens}} pipeline used in this paper are publicly available. The CNN code and simulated images can be made available upon a reasonable request to the corresponding author.

\bibliographystyle{mnras}
\bibliography{Citations}
\label{sec:bib}

\appendix

\section{Our conventions for Einstein radius and Position Angle}
\label{app:conv}

We convert the Einstein radius values quoted in S19 and S23b to our convention \citep[GRAVLENS, ][]{gravlens} using the following relation :
\begin{align}
\teOur = \sqrt{\frac{2q}{1+q^2}}\teAle=\sqrt{\frac{2q}{1+q^2}}\teYL=\sqrt{\frac{2q}{1+q^2}}\frac{2\sqrt{q}}{(1+q)}\teHS
\end{align}
where, $q$ is the ratio of semi-minor to semi-major axis of the mass distribution of the SIE.
We measure the position angle East of North as shown in the Fig.~\ref{fig:10common} and convert the position angle values quoted in S19 and S23b to this convention before comparing. These conversions were derived by equating the form of the convergence assumed in each of these methods. We further obtain the two components of ellipticity using the axis ratio ($q$) and the position angle ($\theta$) in the following relations :
\begin{align}
 e_x=\frac{1-q^2}{1+q^2} \cos (2 \theta) \\ 
 e_y=\frac{1-q^2}{1+q^2} \sin (2 \theta) 
\end{align}

\section{Errors on the network predictions}
\label{app:errors}
\begin{figure*}
\begin{center}
\begin{subfigure}[b]{\textwidth}
    \includegraphics[width=0.49\textwidth]{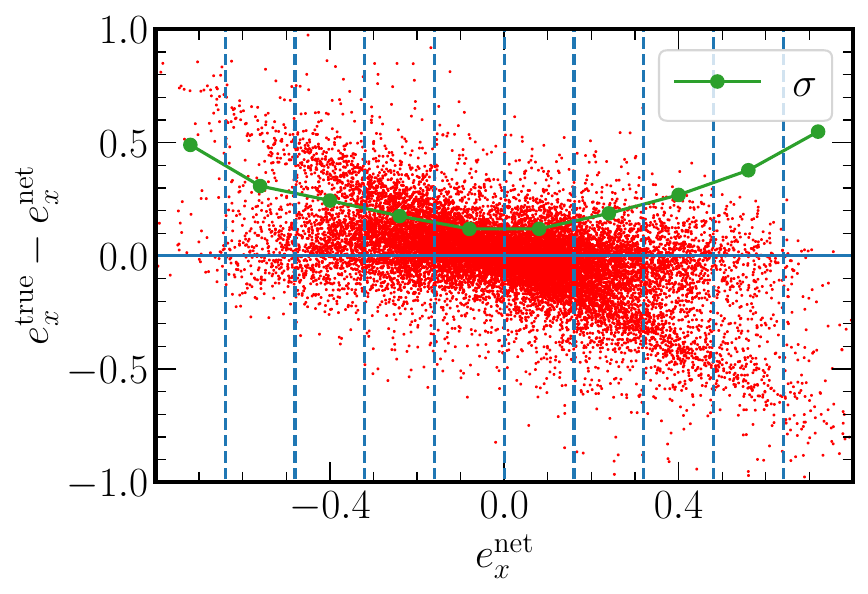} 
    \includegraphics [width=0.49\textwidth]{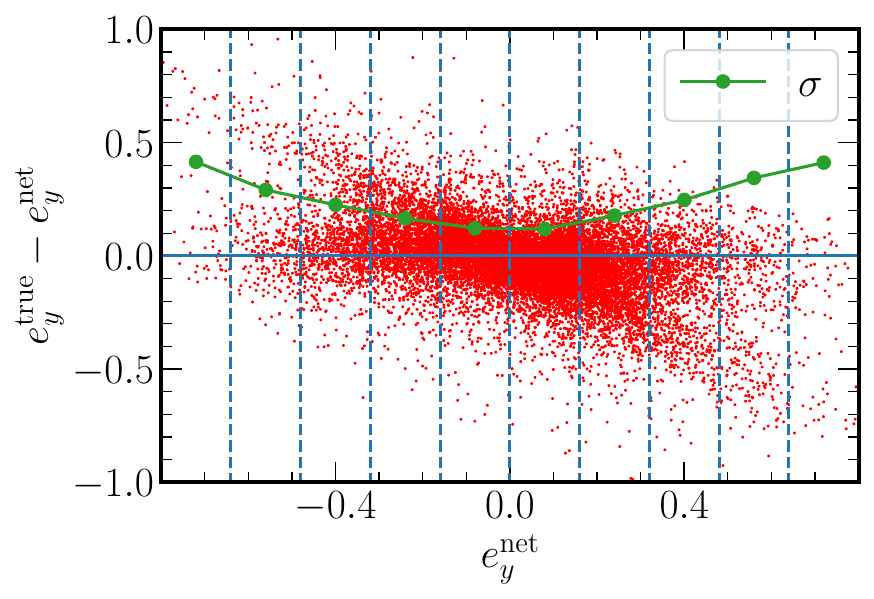}  
\end{subfigure}

\includegraphics[scale=0.625]{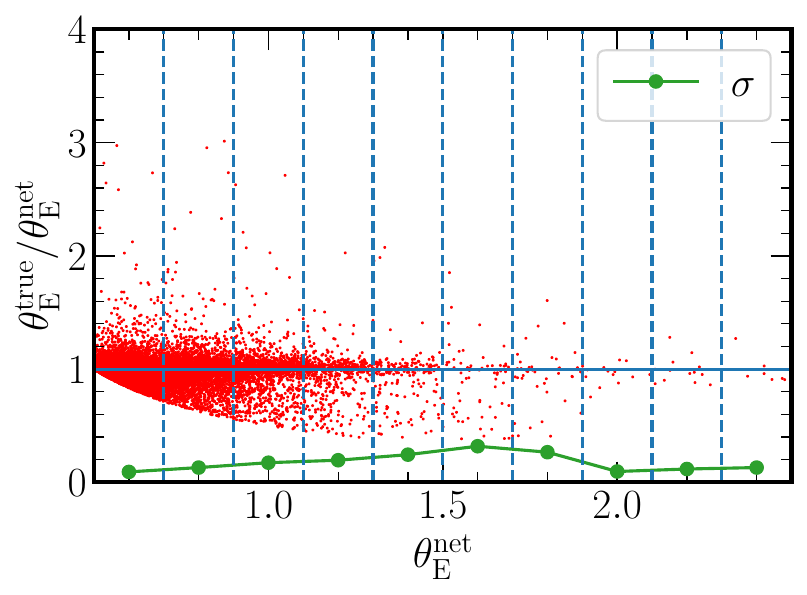} 
\caption{For Einstein radius ($\theta_{\rm E}$), the red points represent a scatter of the ratio of the true value and the network predicted value versus the network value, we divide the parameter range shown in $10$ bins and calculate the standard deviation in each bin as shown by the green points. For the ellipticity components ($e_{x}$ and $e_{y}$), the red points represent a scatter of the difference of the true value and the network value versus the network value, we divide the parameter range in $10$ bins similarly and calculate the standard deviation in each bin as shown by the green points.}
\label{fig:errors}\end{center} 
\end{figure*}

We quantify the error in our network predictions based on the performance of the network on the simulated test sample shown in Fig.~\ref{fig:HSCempty}. For example, we present a scatter plot of the ratio of the true Einstein radius to the network-predicted value as a function of the latter in Fig.~\ref{fig:errors}. We divide the sample in $10$ bins and calculate the standard deviation of the afore-mentioned ratio in each bin (green solid curve). For a given SuGOHI grade A lens, the network prediction is then multiplied by this standard deviation, in the corresponding bin, to obtain the error on the network-predicted Einstein radius in Fig.~\ref{fig:10common}. We follow a similar procedure for assigning errors on $e_{x}$ and $e_{y}$, except by using the scatter in the difference of the true value and the network value as shown in Fig.~\ref{fig:errors}. The computed standard deviation for $e_{x}$ and $e_{y}$ is used as an error from the network in the corresponding bin in tFig.~\ref{fig:10common}.

\section{Performance of our LensLight sample on the grade A SuGOHI lenses}
\label{app:gradeA}
\begin{figure*}
\begin{center}
\includegraphics[scale=.45]{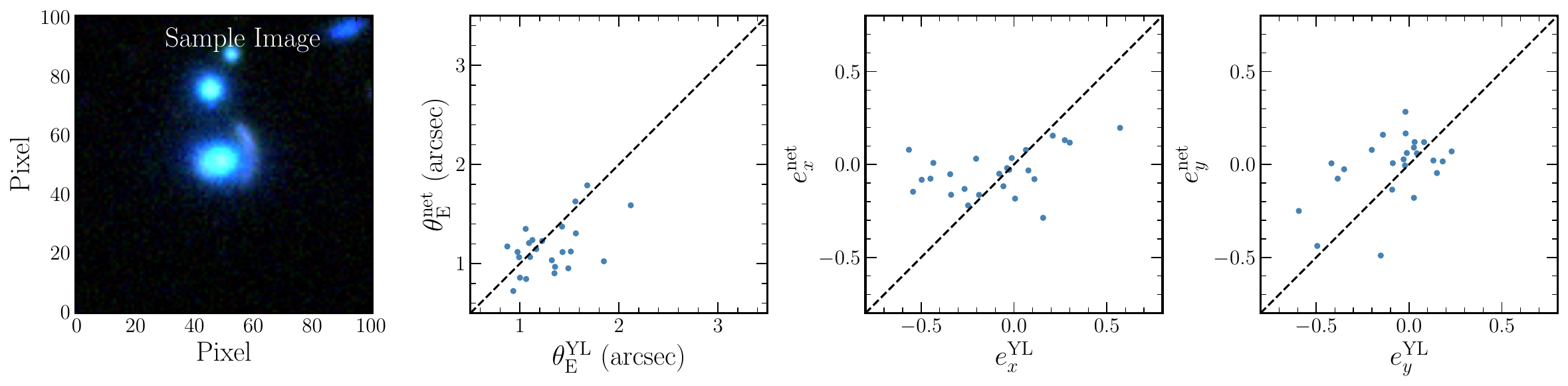}
\caption{Performance of our network (trained with the LensLight sample) on a sample of 25 SuGOHI grade A lenses (not processed by \yatta). The range of the parameters is kept similar to the fig.4 of S23b for ease of comparison.}\label{fig:gradeA}
\end{center} 
\end{figure*}

 We show the performance of our LensLight sample on 182 SuGOHI (grade A and B) lenses in the lower panel of the Fig.~\ref{fig:LensLight}. In S23b, they train their ResNet on a sample qualitatively similar to our LensLight sample and test it on a sample of 31 grade A SuGOHI lenses. In order to compare with S23b, here we show the performance of our LensLight sample on a sample 25 grade A suGOHI lenses (see Fig.~\ref{fig:gradeA}). Our grade A sample is smaller than the S23b sample, since \yatta does not identify a few of these 31 grade A lenses as lenses and we do not have the parameters from \yatta to compare our results with for those lenses. If we compare the results shown in the Fig.~\ref{fig:gradeA} (we compare our network predictions with \yatta) with the fig.4 in S23b (they compare their network predictions with GLEE \& GLAD), we can see that our results are comparable with theirs for the Einstein radius and the components of the ellipticity.

\section{Lens model parameters for SuGOHI lenses and Comparison with \yatta}
\label{app:182AB}
The performance on 182 SuGOHI (grade A and B) lenses of our network trained on the HSCempty sample is shown in the lower panel of the Fig.~\ref{fig:HSCempty}. Here, we present the lens model parameters and errors for these SuGOHI lenses as estimated by \yatta and our network (trained on the HSCempty sample). We only show 25 grade A lenses in the Table~\ref{tab:182_table}. The lens model parameters for 157 grade B lenses, along with their errors are provided in the supplementary material.

\begin{figure*}
\begin{center} 

\includegraphics[scale=0.4]{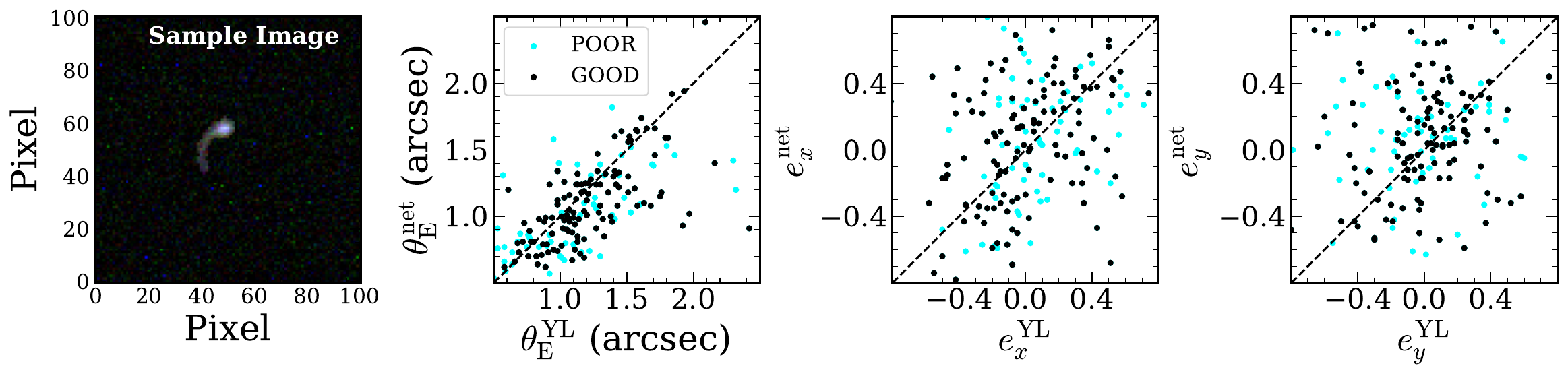}
\includegraphics[scale=0.4]{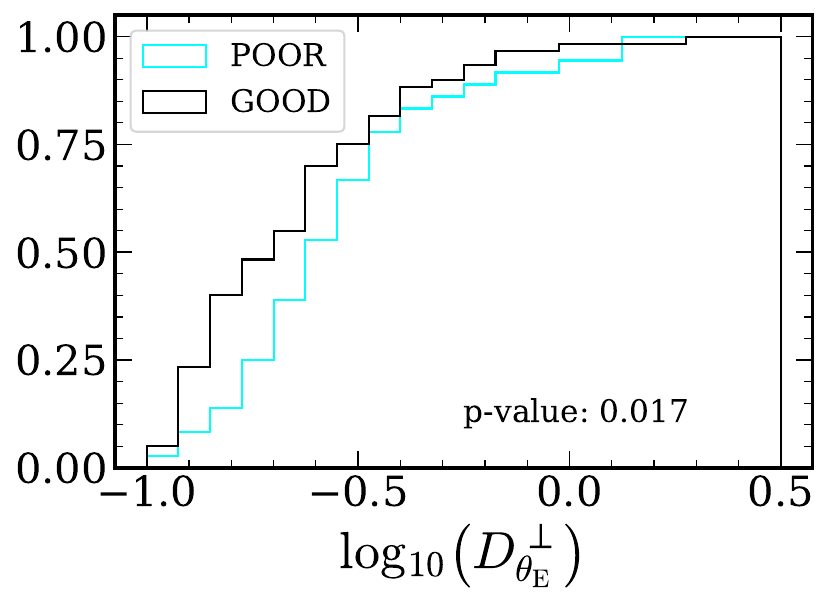}
\includegraphics[scale=0.4]{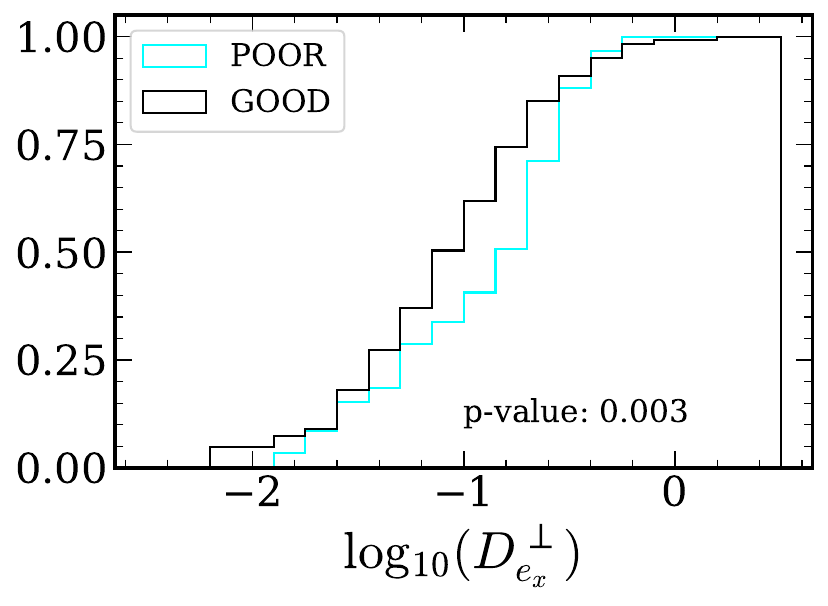}
\includegraphics[scale=0.4]{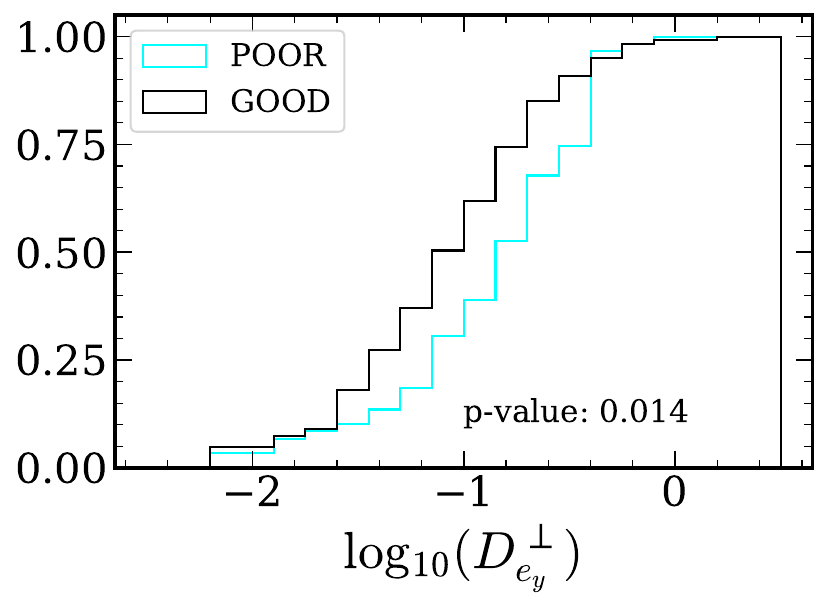}
\caption{Upper panel: Performance of our network trained on the HSCempty sample, evaluated on a test set of 182 SuGOHI (grade A+B) lenses. These lenses were processed using \yatta for central lens light subtraction. Based on visual inspection of the output of YL, we classify the lenses into two categories, GOOD: lenses where \yatta performs both lens light subtraction and modelling successfully (122 lenses, shown in black) and POOR: lenses where either component fails (60 lenses, shown in cyan).\\
Lower panel: Normalised cumulative distributions of the perpendicular distances from the $x = y$ line for the three lens parameters for the GOOD and POOR samples. We perform a Kolmogorov–Smirnov (KS) test to assess whether the two distributions differ significantly, and report the corresponding $p$-value.}
\label{fig:ks}
\end{center}
\end{figure*}

To evaluate the reliability of our network predictions on real data, we evaluate its performance on a test sample of 182 SuGOHI lenses (grade A + B), processed using \yatta for central lens light subtraction and mass modeling. Based on visual inspection of the \yatta output, we classify the lenses into two categories: GOOD, comprising 122 lenses where both the lens light subtraction and mass modelling appeared to perform well; and POOR, comprising 60 lenses where one or both steps appeared suboptimal as shown in the upper panel of Fig.~\ref{fig:ks}.

To explore whether the networks predictions correlate differently with the \yatta estimates across these categories, we calculate the perpendicular distances from the $x = y$ line in parameter comparison plots and analyse the normalised distributions of their logarithms for GOOD and POOR lenses. A Kolmogorov–Smirnov (KS) test reveals statistically significant differences between the GOOD and POOR distributions, with $p$-values of 0.017 for $\theta_{\rm E}$, 0.003 for $e_x$, and 0.014 for $e_y$ (see lower panel of Fig.~\ref{fig:ks}). These results indicate that the network predictions align more closely with \yatta estimates for the lenses where the lens light subtraction and mass modelling by \yatta is visually assessed to be more reliable.

The average time required for \yatta to perform a quick lens light subtraction is approximately 15 s per lens, with more refined procedures taking even longer. Given the large volume of lenses expected from upcoming surveys such as LSST, this presents a computational challenge. To address this, there is a need to develop faster lens light subtraction techniques, potentially leveraging machine learning approaches such as convolutional autoencoders like U-Net architectures, to enable scalable preprocessing at survey scale. Developing such an algorithm would require substantial effort and is therefore beyond the scope of this work. However, it represents a promising direction for future research.

\begin{table}
	\centering
	\caption{Lens model parameters predicted by our network (CNN) and by \yatta, along with errors, for 182 SuGOHI grade A and B lenses (we only show grade A lenses here and grade B lenses are provided in the supplementary material.)}
	
     \label{tab:182_table}
    \begin{adjustbox}{width=\columnwidth,keepaspectratio}
    \begin{tabular}{llccr p{10cm}}
        \hline
        Name  & Method & $\theta_{\rm E}$ (arcsec) & $e_{x}$ & $e_{y}$ \\
        \hline
HSCJ015618-010747 & CNN & $0.80 \pm 0.11$ & $-0.65 \pm 0.49$ & $-0.44 \pm 0.23$\\
 & \yatta & $0.87 \pm 0.18$ & $-0.34 \pm 0.26$ & $-0.35 \pm 0.23$\\
\hline
HSCJ020141-030946 & CNN & $1.56 \pm 0.50$ & $0.29 \pm 0.19$ & $0.02 \pm 0.12$\\
 & \yatta & $1.49 \pm 0.12$ & $0.27 \pm 0.14$ & $0.18 \pm 0.15$\\
\hline
HSCJ020241-064611 & CNN & $1.38 \pm 0.34$ & $0.08 \pm 0.12$ & $0.14 \pm 0.12$\\
 & \yatta & $1.06 \pm 0.42$ & $-0.20 \pm 0.21$ & $-0.20 \pm 0.19$\\
\hline
HSCJ020955-024442 & CNN & $0.80 \pm 0.10$ & $-0.20 \pm 0.18$ & $0.47 \pm 0.25$\\
 & \yatta & $0.93 \pm 0.17$ & $0.06 \pm 0.24$ & $-0.09 \pm 0.21$\\
\hline
HSCJ021737-051329 & CNN & $0.79 \pm 0.10$ & $-0.76 \pm 0.49$ & $-0.11 \pm 0.12$\\
 & \yatta & $1.17 \pm 0.11$ & $-0.45 \pm 0.09$ & $-0.14 \pm 0.15$\\
\hline
HSCJ022346-053418 & CNN & $1.20 \pm 0.24$ & $0.34 \pm 0.27$ & $0.45 \pm 0.25$\\
 & \yatta & $0.98 \pm 0.29$ & $-0.54 \pm 0.25$ & $0.08 \pm 0.26$\\
\hline
HSCJ022610-042011 & CNN & $1.12 \pm 0.22$ & $0.55 \pm 0.38$ & $0.02 \pm 0.12$\\
 & \yatta & $1.10 \pm 0.17$ & $-0.08 \pm 0.21$ & $-0.02 \pm 0.21$\\
\hline
HSCJ023217-021703 & CNN & $1.03 \pm 0.18$ & $0.09 \pm 0.12$ & $0.68 \pm 0.41$\\
 & \yatta & $1.35 \pm 0.35$ & $-0.01 \pm 0.23$ & $-0.42 \pm 0.23$\\
\hline
HSCJ023322-020530 & CNN & $0.97 \pm 0.17$ & $0.41 \pm 0.27$ & $0.00 \pm 0.12$\\
 & \yatta & $1.56 \pm 0.08$ & $0.01 \pm 0.10$ & $-0.01 \pm 0.09$\\
\hline
HSCJ085046+003905 & CNN & $1.15 \pm 0.23$ & $0.47 \pm 0.27$ & $0.72 \pm 0.41$\\
 & \yatta & $1.57 \pm 0.08$ & $0.57 \pm 0.08$ & $-0.09 \pm 0.07$\\
\hline
HSCJ085855-010208 & CNN & $1.08 \pm 0.19$ & $-0.11 \pm 0.12$ & $0.01 \pm 0.12$\\
 & \yatta & $1.06 \pm 0.04$ & $-0.03 \pm 0.07$ & $0.03 \pm 0.06$\\
\hline
HSCJ090429-010227 & CNN & $0.79 \pm 0.10$ & $0.25 \pm 0.19$ & $-0.11 \pm 0.12$\\
 & \yatta & $0.40 \pm 0.13$ & $0.16 \pm 0.33$ & $-0.82 \pm 0.24$\\
\hline
HSCJ121052-011905 & CNN & $1.34 \pm 0.33$ & $0.26 \pm 0.19$ & $-0.07 \pm 0.12$\\
 & \yatta & $1.23 \pm 0.11$ & $0.21 \pm 0.17$ & $-0.38 \pm 0.15$\\
\hline
HSCJ124320-004517 & CNN & $1.61 \pm 0.52$ & $-0.21 \pm 0.18$ & $-0.17 \pm 0.17$\\
 & \yatta & $1.51 \pm 0.23$ & $-0.27 \pm 0.18$ & $0.04 \pm 0.16$\\
\hline
HSCJ125254+004356 & CNN & $1.46 \pm 0.36$ & $0.31 \pm 0.19$ & $-0.26 \pm 0.17$\\
 & \yatta & $1.13 \pm 0.11$ & $-0.57 \pm 0.17$ & $0.03 \pm 0.25$\\
\hline
HSCJ135138+002840 & CNN & $2.41 \pm 0.32$ & $0.31 \pm 0.19$ & $-0.03 \pm 0.12$\\
 & \yatta & $2.12 \pm 0.06$ & $0.11 \pm 0.11$ & $0.23 \pm 0.20$\\
\hline
HSCJ141136-010215 & CNN & $0.99 \pm 0.17$ & $-0.29 \pm 0.18$ & $-0.00 \pm 0.12$\\
 & \yatta & $1.00 \pm 0.05$ & $-0.06 \pm 0.10$ & $-0.49 \pm 0.15$\\
\hline
HSCJ142720+001916 & CNN & $1.36 \pm 0.33$ & $0.10 \pm 0.12$ & $0.36 \pm 0.25$\\
 & \yatta & $1.36 \pm 0.13$ & $-0.04 \pm 0.17$ & $0.13 \pm 0.17$\\
\hline
HSCJ144320-012537 & CNN & $0.92 \pm 0.16$ & $0.28 \pm 0.19$ & $0.59 \pm 0.34$\\
 & \yatta & $1.09 \pm 0.06$ & $-0.19 \pm 0.10$ & $-0.59 \pm 0.13$\\
\hline
HSCJ145242+425732 & CNN & $0.98 \pm 0.17$ & $0.25 \pm 0.19$ & $-0.62 \pm 0.29$\\
 & \yatta & $1.85 \pm 0.45$ & $-0.25 \pm 0.25$ & $-0.02 \pm 0.26$\\
\hline
HSCJ223733+005015 & CNN & $1.51 \pm 0.48$ & $-0.14 \pm 0.12$ & $-0.08 \pm 0.12$\\
 & \yatta & $1.32 \pm 0.23$ & $0.08 \pm 0.18$ & $0.15 \pm 0.17$\\
\hline
HSCJ230335+003703 & CNN & $0.83 \pm 0.11$ & $-0.44 \pm 0.24$ & $-0.64 \pm 0.42$\\
 & \yatta & $0.99 \pm 0.27$ & $-0.50 \pm 0.28$ & $-0.02 \pm 0.21$\\
\hline
HSCJ230521-000211 & CNN & $1.06 \pm 0.19$ & $0.20 \pm 0.19$ & $-0.61 \pm 0.29$\\
 & \yatta & $1.68 \pm 0.02$ & $-0.34 \pm 0.06$ & $-0.03 \pm 0.02$\\
\hline
HSCJ233130+003733 & CNN & $1.35 \pm 0.33$ & $0.15 \pm 0.12$ & $-0.15 \pm 0.12$\\
 & \yatta & $1.43 \pm 0.06$ & $0.30 \pm 0.07$ & $-0.15 \pm 0.05$\\
\hline
HSCJ233146+013845 & CNN & $1.06 \pm 0.18$ & $-0.80 \pm 0.49$ & $-0.19 \pm 0.17$\\
 & \yatta & $1.43 \pm 0.17$ & $-0.43 \pm 0.12$ & $0.03 \pm 0.13$\\
\hline

    \end{tabular}
    \end{adjustbox}
\end{table}

\bsp
\label{lastpage}
\end{document}